\documentclass[12pt]{article}
\pdfoutput=1
\usepackage{jheppub}

\usepackage{amsfonts}
\usepackage{amscd}
\usepackage{amssymb}
\usepackage{amsmath,bbm}
\usepackage{epsfig}
\usepackage{latexsym}
\usepackage{mathtools}
\usepackage{arydshln}
\usepackage{hyperref}
\usepackage{color}
\usepackage{slashed}
\usepackage{euscript, mathrsfs}
\usepackage[vcentermath]{youngtab}
\usepackage{marginnote}
\usepackage{bm}
\usepackage{comment}
\usepackage{multirow}

\def\be{\begin{equation}}
\def\ee{\end{equation}}
\def\bea{\begin{eqnarray}}
\def\eea{\end{eqnarray}}
\def \beaa {\begin{equation}\begin{aligned}}
\def \eeaa {\end{aligned}\end{equation}}

\newcommand{\nn}{\nonumber}

\newcommand\diff{\mathrm{d}}

\newcommand{\ii}{\mathrm{i}}

\newcommand{\vol}{\mathrm{vol}}

\newlength{\sswidth}

\def \be  {\begin{equation}}
\def \ee  {\end{equation}}
\def \ba  {\begin{eqnarray}}
\def \ea  {\end{eqnarray}}
\def \bb  {\begin{thebibliography}}
\def \eb  {\end{thebibliography}}
\def \lab #1 {\label{#1}}

\newcommand\cD{\mathcal{D}}
\newcommand\cF{\mathcal{F}}

\newcommand\cI{\mathcal{I}}

\newcommand\cK{\mathcal{K}}

\newcommand\cM{\mathcal{M}}
\newcommand\cN{\mathcal{N}}
\newcommand\cO{\mathcal{O}}

\newcommand\cU{\mathcal{U}}

\newcommand\bo{\bar{1}}
\newcommand\bt{\bar{2}}

\newcommand\tr{\mathrm{Tr}}

\newcommand\ie{\textit{i.e.}}

\newcommand{\diag}{{\rm diag}\,}

\title{K{\"a}hler Uniformization from Holographic Renormalization Group Flows of M5-branes}
\author{Martin Fluder}
\affiliation{
Walter Burke Institute for Theoretical Physics,
California Institute of Technology\\
Pasadena, CA 91125, USA
}

\emailAdd{fluder@caltech.edu}

\preprint{CALT-TH-2017-038}

\abstract{
In this paper, we initiate the study of holographic renormalization group flows for the metric of four-manifolds. In particular, we derive a set of equations which govern the evolution of a generic K\"ahler four-manifold along the renormalization group flow in seven-dimensional gauged supergravity. The physical eleven-dimensional M-theory setup is given by a stack of M5-branes wrapping a calibrated K\"ahler four-cycle inside a Calabi-Yau threefold. By topologically twisting the theory in the ultraviolet, we may choose an arbitrary K\"ahler metric on the four-cycle as an asymptotic boundary condition. We find that at the infrared fixed point, we reach a K\"ahler-Einstein metric, which can be interpreted as an indication of ``uniformizing" behavior of the flow.
}

\begin{document}

\maketitle

\bibliographystyle{JHEP}

\section{Introduction}\label{SecIntroduction}


The study of smooth four-manifolds is a rich and still largely puzzling subject in mathematics. For instance, it is presently unknown how to classify simply connected compact smooth four-manifolds. This is the case despite the fact that the classification of \emph{topological} four-manifolds has been proved a long time ago~\cite{freedman1982}. To this day, it is an open problem how to translate the classification of topological four-manifolds into a classification of smooth four-manifolds. One issue lies in the fact that there are examples of spaces, such as $\mathbb{R}^{4}$, which have an uncountable number of different smooth structures (known as exotic $\mathbb{R}^{4}$). Similarly, it appears that some important tools, crucial in the study and classification of manifolds in lower dimensions, seem to be less powerful in the case of four-dimensional smooth manifolds. For instance the Ricci flow equation~\cite{hamilton1982,Hamilton1} (a well-known example of a flow, that ``uniformizes" the metric), which was famously employed in the proof of Thurston's geometrization conjecture of three-manifolds~\cite{Thurston} (and in particular the Poincar\'e conjecture) in~\cite{Perelman:2006un}, does not preserve the Hermiticity of the metric.\footnote{There is a variant of the Ricci flow -- the Hermitian Ricci flow -- which does preserve Hermitian metrics along the flow~\cite{2008arXiv0804.4109S} also related to the physics of RG flows~\cite{doi:10.1093/imrn/rnp237, 2010arXiv1008.2794S}, and various interesting results have been proved by the same authors. However, a uniform treatment of smooth four-manifolds using Ricci flows seems to be lacking as of now.} The Ricci flow has a natural interpretation in physics~\cite{Friedan:1980jm}; it arises as the renormalization group flow of the target-manifold of a two-dimensional sigma-model at one-loop. From this (physical) perspective, the fact that solutions to the Ricci flow equation approach constant curvature metrics can be viewed as the consequence of the renormalization group flow ``washing out" (irrelevant) data such as the moduli of the original metric.\footnote{An alternative vastly successful approach to the study of four-manifold motivated (also) from physics is by the use of gauge theory~\cite{Witten:1988ze,Seiberg:1994rs,Seiberg:1994aj,Witten:1994cg}. This is unrelated to the motivation of the current paper} 

This is a common theme when studying renormalization group flows in physics, and a natural question to ask is whether more intricate (physically relevant) setups could give rise to novel ``uniformization" flows that might help in the study of smooth (four-) manifolds.\footnote{Throughout this paper, we shall use the term uniformization to describe metric (renormalization group) flows which we believe to exhibit (loosely speaking) behavior leading to uniform (\emph{e.g.} constant-curvature) metric at the (infrared) fixed point of the flow.} In this paper, we employ this approach, and study the supergravity limit of a stack of M5-branes wrapping a K\"ahler four-manifold in M-theory. This leads to holographic renormalization group flows for the metric of K\"ahler four-manifolds, which we expect (on physical grounds) to be ``uniformizing".

The study of branes wrapping supersymmetric cycles from the perspective of holography was first introduced in~\cite{Maldacena:2000mw}. In particular, they adopted the perspective of viewing these setups as holographic renormalization group flows across dimensions. Their focus was on the case of M5-branes (among other examples) to wrap a Riemann surface. Subsequently, a plethora of solutions describing M5-branes wrapping certain classes of four-manifolds has been found by studying either the effective maximally supersymmetric seven-dimensional gauged supergravity (a consistent truncation of eleven-dimensional supergravity) or the full M-theory supergravity background~\cite{Gauntlett:2000ng,Gauntlett:2001jj,Benini:2012cz,Benini:2013cda,Karndumri:2015sia,Bah:2015nva}.\footnote{In the case of M5-branes wrapping four-manifolds, we detail the relevant setups and allowed classes of four-manifolds involved in Section~\ref{Sec:twisting} (see also Table~\ref{table:4cyclecalb}).} 

In order to preserve some supersymmetry, the theory will generally be required to be ``twisted"~\cite{Witten:1988ze,Bershadsky:1995qy}. Due to the twist, \emph{a priori} any choice of metric on a four-manifold (within a given class) will preserve some supersymmetry. However, most supergravity solutions known thus far assume that the twists hold along the \emph{full} renormalization group flow from the ultraviolet asymptotically locally AdS$_{7}$ to the infrared AdS$_{3}$. This then puts constraints on the particular type of four-manifolds allowed by supersymmetry, because the flow has to be consistent with an AdS$_{3}$ solution in the deep infrared. 
 However, since the metric is fixed along the full flow, one cannot observe how it varies along the flows, and thus the expected uniformization of the metric is not visible. 

In reference~\cite{Anderson:2011cz}, the authors remedy this by working out the case of M5-branes (among other examples) wrapping Riemann surfaces, but now with the metric on the Riemann surface left arbitrary. They prove that indeed flows exist and uniformize the metric on the Riemann surface. This result is motivated from the corresponding field theory setup, which states that upon wrapping M5-branes on a Riemann surface, the resulting four-dimensional $\cN=2$ superconformal field theories should only depend on the complex structure of the curve~\cite{Gaiotto:2009we,Gaiotto:2008cd,Gaiotto:2009hg}.

In this paper, we aim to initiate an extension of the discussion of holographic renormalization group flows across dimensions; we consider a physical setup of M5-branes wrapping K\"ahler four-cycles, which are calibrated cycles inside a Calabi-Yau threefold. The physical setup then requires the twist to be implemented in the ultraviolet as an asymptotic boundary condition, such that supersymmetry is preserved. Similarly, in the infrared, it is required that the solution is in fact a valid (vacuum) AdS$_{3}$ solution, which puts asymptotic constraints on the fields and the metric. As opposed to the solutions in~\cite{Gauntlett:2000ng,Gauntlett:2001jj,Benini:2012cz,Benini:2013cda,Karndumri:2015sia,Bah:2015nva}, both the ultraviolet and the infrared thus merely serve as boundary conditions, and one studies the equations arising from supergravity (and in particular the condition that \emph{some} supersymmetry is preserved along the full flow\footnote{In this paper, we shall restrict to flows that are ${1\over 2}$-BPS with respect to the maximally possible supersymmetry allowed for such a setup.}) for the metric of the four-manifold in the bulk of the flow.\footnote{As remarked in~\cite{Anderson:2011cz}, one views such types of supergravity flows as a boundary-value problem, with prescribed infrared and ultraviolet boundary conditions. However, this is rather different to the usual picture of Wilsonian renormalization group flows.} 

Apart from the supersymmetry (and their integrability) conditions we solve all equations of motions, Einstein equations, and Bianchi identities for the metric of the K\"ahler four-manifold. We find a set of equations for the metric, which, if they are satisfied, give a solution to the full supergravity setup. They are second-order in terms of the K\"ahler metric.  We further study the boundary conditions in the ultraviolet as well as the infrared. We find that indeed in the ultraviolet (to leading order) there will not be any constraints on the K\"ahler metric. At the infrared fixed point however, we observe that the supergravity equations imply that the metric has to be K\"ahler-Einstein. This can be viewed as an indication of uniformizing behavior of the set of equations we derive for the K\"ahler metric. 

This paper is organized as follows. We start in Section~\ref{Sec:twisting} by reviewing some aspects of twisted M5-branes and their relation to calibrated cycles of special holonomy manifolds. We further introduce some intuition behind the notion of uniformization (or its higher dimensional analogue) arising when wrapping M5-branes on calibrated four-cycles. In Section~\ref{Sec:7dGauged} we introduce our main tool, namely the maximally supersymmetric seven-dimensional gauged supergravity. In Section~\ref{Sec:Ansatz}, we discuss our ansatz and provide some more details for the particular calibration considered in this paper. Finally, in Section~\ref{Sec:Solution}, we present the metric flow equations and make some comments about their asymptotic behavior. Lastly, in Section~\ref{Sec:Discussion}, we conclude the main part of the paper with a discussion of our results and a rather extensive list of interesting future directions. In two appendices we provide some clarification of our notation in the main part of the paper, and some more details for the derivation of our solution.


\section{Twisting, calibrated cycles and uniformization}
\label{Sec:twisting}


Generically, when putting a supersymmetric theory on a curved manifold, we will not be able to preserve (any) supersymmetry. This is due to the fact that there might not exist a covariantly constant Killing spinor. However, if the theory has global symmetries (such as R-symmetries or flavor symmetries) one can implement what is called a (partial) topological twist~\cite{Witten:1988ze}. The idea is that one introduces a background field $A_{\mathrm{g}}$ for (part of) the global symmetry group and tunes it in such a way that it cancels against (part of) the spin-connection, \ie~somewhat schematically
\bea\label{Eqn:schemtwist}
\partial_{\mu} \epsilon +\left(  \omega_{\mu}{}^{ab} \gamma_{ab}  + A_{\mathrm{g}} \right) \epsilon \ = \ \partial_{\mu} \epsilon \ = \ 0 \,.
\eea 
In principle one may choose any part of the global symmetry group to perform this twist. However, since one would prefer this procedure to be independent of the choice of theory, it is advantageous to use part of the global R-symmetry group~\cite{Bobev:2017uzs}. Since the stress energy tensor is in the same supermultiplet as the R-current, there will always be a way to twist, independent of the details of the theory.

For the purpose of this paper we shall be interested in branes wrapping (arbitrary) supersymmetric cycles. It is then generically the case that the field theory realized on the branes is twisted~\cite{Bershadsky:1995qy}. In particular, the cycles will typically not have any covariantly constant spinors, and hence supersymmetry has to be preserved by implementing a (partial) topological twist. 

An alternative point of view on such twists is to start with the full (string or) M-theory. In order to preserve supersymmetry in the full eleven-dimensional M-theory setup, we have to put the theory on a ``special holonomy manifold". Then, to support static M5-branes solutions, we require the M5-branes to wrap supersymmetric cycles of the special holonomy manifold. It can be argued on general grounds that these supersymmetric cycles are precisely given by calibrated cycles~\cite{Becker:1995kb,Becker:1996ay,Gauntlett:1998vk,Gibbons:1998hm} (see also~\cite{Gauntlett:2003di} for a nice review).\footnote{We call a $q$-form $\Phi$ on a manifold $M$ a calibration if and only if $\diff \Phi = 0$, and $\forall x \in M$ and any oriented $q$-dimensional subspace $\xi_{x} \subset T_{x}M$, $\left. \Phi \right|_{\xi_{x}} \leq \left. \vol \right|_{\xi_{x}}$, where $\left. \vol \right|_{\xi_{x}}$ is the volume form of $\xi_{x}$. A $q$-cycle $\Sigma_{q}$ is then calibrated by $\Phi$ if and only if 
\bea
 \left. \Phi \right|_{\Sigma_{q}} \ \equiv \ \left. \vol \right|_{\Sigma_q} \,.
\eea}

In the current paper, we are mainly concerned with the case of calibrated \emph{four}-cycles. In Table~\ref{table:4cyclecalb}, we detail the possible calibrated four-cycles of M-theory on special holonomy manifolds, as well as the preserved supersymmetry in two dimensions, and the condition on the four-cycles arising from explicit supergravity solutions. All but one of these calibrated brane setups have a corresponding global solution in a truncated gauged seven-dimensional supergravity~\cite{Gauntlett:2000ng,Gauntlett:2001jj,Benini:2012cz,Benini:2013cda}.\footnote{Any solution in the truncated gauged seven-dimensional supergravity can be uplifted to eleven-dimensional M-theory, as we will discuss in some detail in Section~\ref{Sec:Uplift}.} The (single) case without a known solution in the effective seven-dimensional setup (\ie~K\"ahler four-cycles in CY$_{3}$) will be treated in this paper.\footnote{In the current paper, we shall not provide or investigate global solutions, since the focus is on deriving the flow equations from supergravity. We intend to study possible global solutions in future work.} However this case has a solution in eleven-dimensional M-theory given by AdS$_{3}\times \text{CY}_{3} \times S^{2}$~\cite{Maldacena:1997re} (see also~\cite{Gauntlett:2006ux,Figueras:2007cn}).
{\renewcommand{\arraystretch}{1.3}
\begin{table}[h!]
\begin{centering}
\begin{tabular}{ | c | c | c | c |}
\hline
 Calibration	& Embedding 					& \  \ $2d$ SUSY \  \ 	& IR manifold $\cM^{(IR)}_{4}$  \\
\hline
\hline
 \multirow{2}{*}{SLAG} &  {$\cM_{4} \subset \text{CY}_{4}$}  & {$\cN=(1,1)$} & \multirow{2}{*}{Constant curvature} \\
  \cdashline{2-3}
  & {$\cM_{2}\times \widetilde{\cM}_{2} \subset \text{CY}_{2}\times \widetilde{\text{CY}}_{2}$} 	& {$\cN=(2,2)$} &\\
  \hline
 \multirow{2}{*}{K\"ahler} 		& $\cM_{4} \subset \text{CY}_{3}$ 	& $\cN=(4,0)$ 		&  \multirow{2}{*}{K\"ahler-Einstein} \\
  \cdashline{2-3}
  			& $\cM_{4} \subset \text{CY}_{4}$ 	& $\cN=(2,0)$ 		&  \\
\hline
&&&K\"ahler-Einstein and \\ 
Lagrangian	& $\cM_{4} \subset \text{HK}_{2}$ 	& $\cN=(2,1)$ 		&   constant holomorphic \\
&&&  sectional curvature\\
\hline
Coassociative & $\cM_{4} \subset G_{2}$ & $\cN=(2,0)$ &   \multirow{2}{*}{Conformally half-flat} \\
\cline{1-3}
Cayley		& $\cM_{4} \subset \mathrm{Spin}(7)$ & $\cN=(1,0)$ & \\
\hline
\end{tabular}
\caption{The possible calibrated four-cycles of special holonomy manifolds (coming from bilinears of spinors). In the first two columns we list the type of calibration and the particular embedding into the special holonomy manifold. In the third column we write down the maximal supersymmetry preserved in the two-dimensional superconformal field theory from the respective calibration (or from the equivalent partial topological twist). Finally, in the fourth column we list the conditions arising from (known) supergravity solutions~\cite{Gauntlett:2000ng,Gauntlett:2001jj,Benini:2012cz,Benini:2013cda} on the four-cycles in the infrared limit, where the geometry is given by AdS$_{3}\times \cM_{4}^{(IR)}$.
} \label{table:4cyclecalb}
\end{centering}
\end{table}}

An alternative perspective on such brane setups is as renormalization group flows across dimensions. On the field theory side, in the ultraviolet of the RG flow, we expect the microscopic description to be given by the worldvolume theory on the M5-branes -- namely the six-dimensional $\cN=(2,0)$ superconformal field theory~\cite{Witten:1995zh,Seiberg:1996vs} -- on some nontrivial background of the form $\mathbb{R}^{1,1} \times \cM_{4}$. Moving to the infrared regime, we expect the characteristic size of $\cM_{4}$ to be small and the theory to be essentially given by a two-dimensional superconformal field theory with the amount of supersymmetry determined by the particular class of four-manifolds, and the twist/choice of calibration (see Table~\ref{table:4cyclecalb}).

Complementary to this field theoretic picture, there is a corresponding dual holographic RG flow analogue. The $\cN=(2,0)$ theory in the ultraviolet is dual to eleven-dimensional supergravity on a space of the form AdS$_{7}\times S^{4}$~\cite{Maldacena:1997re, Berkooz:1998bx}. However, to match the field theory setup, the AdS$_{7}$-factor is now given by $\mathbb{R}^{1,1} \times \cM_{4}$ at constant $r$-slices, with $r$ the radial direction of AdS$_{7}$. In order be able to put the theory on such a background, we have to precisely implement the (partial) topological twist in the ultraviolet, given schematically by the condition in~\eqref{Eqn:schemtwist}. In the infrared however, we expect a two-dimensional superconformal field theory and therefore the corresponding supergravity dual should be of the form AdS$_{3}\times \cM_{4}$ with a certain amount of supersymmetry preserved.

A priori, the internal four-manifold $\cM_{4}$ can be picked arbitrarily within a given class of calibrations. Due to the (partial) twist, supersymmetry is guaranteed to be preserved. However, as was observed in the particular supergravity solutions~\cite{Gauntlett:2000ng,Gauntlett:2001jj,Benini:2012cz,Benini:2013cda}, supersymmetry imposes further conditions if we want solutions which also exist in the deep infrared and give an appropriate physically relevant AdS$_{3}$ solution. In all of those cases the internal four-cycle $\cM_{4}$ and in particular the twisting condition in~\eqref{Eqn:schemtwist} was \emph{fixed} along the full flow from the ultraviolet to the infrared. 

In the current paper, we are precisely interested in studying how the metric varies along the RG flow. This was considered in~\cite{Anderson:2011cz} for the calibrated cycle given by a Riemann surface. In the following, we shall employ a similar strategy but for four-cycles. If we leave the metric arbitrary, the infrared and ultraviolet behavior become separate asymptotic boundary conditions to a set of equations which determines the RG flow of the metric. The (partial) topological twist is then only applicable in the ultraviolet, and so we may pick \emph{any} choice of four-cycle (within a given class of calibrations) asymptotically in the UV. In the asymptotic infrared region however, we generally expect to reach the known AdS$_{3}$ solutions and consequently we expect that the additional conditions on the IR four-cycles have to be satisfied. In the following to distinguish the two asymptotic metrics, we shall denote the four-manifold appearing in the ultraviolet as $\cM_{4}^{(UV)}$, and the one in the infrared as $\cM_{4}^{(IR)}$; of course they are still the ``same" manifold, but with different metrics on it. In Table~\ref{table:4cyclecalb}, we provide the expected infrared conditions for $\cM_{4}^{(IR)}$ arising from the known holographic solutions.

The reason one expects further conditions on the four-cycle in the infrared regime, can be understood by considering again the dual field theory setup; we take the field theory limit on the stack of M5-branes, and expect to flow to a two-dimensional field theory. Far in the infrared regime, it is expected that we obtain a conformal fixed point, which is precisely the theory dual to the IR AdS$_{3}$ solution. The precise details of the metric on the four-cycle $\cM_{4}^{(IR)}$ then enter as data for the ``effective" superconformal field theory at the fixed point. For the case of four-dimensional $\cN=2$ superconformal field theories arising on M5-branes wrapping calibrated two-cycles inside CY$_{2}$, only the complex structure of the Riemann surface enters the description of the four-dimensional theory~\cite{Gaiotto:2009we,Gaiotto:2008cd,Gaiotto:2009hg}. At the same time, the conformal factor of the metric is supposed to be washed out along the RG flow. Therefore, it is expected that the internal Riemann surface ``uniformizes" along the RG flow. This is precisely the uniformization behavior observed in~\cite{Anderson:2011cz} from holography.

Similar results are expected to hold for the case of M5-branes wrapping higher-dimensional calibrated cycles~\cite{Dimofte:2011ju,Dimofte:2011py,Gadde:2013sca}.\footnote{See also some discussion in the introduction of~\cite{Dimofte:2014ija}.} Thus, we expect that the infrared four-cycle $\cM_{4}^{(IR)}$ ``uniformizes" in the deep infrared, which is what we set out to test in the following using holography.


\section{Seven-dimensional maximally supersymmetric gauged supergravity}\label{Sec:7dGauged}


In this section, we set up the maximally supersymmetric seven-dimensional $SO(5)$ gauged supergravity theory as introduced in~\cite{Pernici:1984xx}. This is the theory in which we are computing the relevant M5-branes renormalization group flows for K\"ahler four-manifolds. The theory has $\cN=4$ supersymmetry and can be obtained by a consistent truncation of eleven-dimensional M-theory on $S^{4}$~\cite{Nastase:1999cb,Nastase:1999kf,Cvetic:1999xp}. As such it has an $SO(5)_g$ gauge symmetry. Furthermore, there is a composite $SO(5)_{c}$ symmetry acting on the scalars.

Apart from the seven-dimensional graviton $e_{\mu}{}^{m}$, the bosonic field content of this theory includes fourteen scalar fields which we package into a tensor $\Pi_{I}{}^{i}$ transforming in the fundamental representation of $SO(5)_g$ from the left and in the fundamental representation of $SO(5)_c$ from the right.\footnote{For our explicit choice of notation and indices, we refer to Appendix~\ref{App:Notation}.} For ease of notation we shall also introduce the fields recombined into a symmetric matrix $T_{ij}$ as follows\footnote{For convenience, we shall switch back and forth between the two notations, $\Pi_{I}{}^{i}$ and $T_{ij}$.}
\bea
T_{ij} \ = \ \left( \Pi^{-1} \right)_i{}^{I} \left( \Pi^{-1} \right)_j{}^{J} \delta_{IJ} \,,\qquad
T \ = \ \delta^{ij} T_{ij}\,,
\eea
which parametrizes the $SL(5,\mathbb{R})/SO(5)_c$ coset and satisfies $\left|\det \left( T_{ij} \right) \right|= 1$. Furthermore, there is a 1-form gauge field $A_{\mu}{}^{IJ}$ transforming in the adjoint of $SO(5)_{g}$ with field strength
\bea
F_{\mu \nu}{}^{IJ} 
 \ \equiv \
\diff A^{IJ} + g A^{I}{}^{K} \wedge A_{K}{}^{J} \,,
\eea
where we denoted by $g$ the seven-dimensional gauge coupling. Similarly, we may introduce symmetric and anti-symmetric composite gauge fields $P_{\mu \, ij}$ and $Q_{\mu \, ij}$ via
\bea
 Q_{\mu\, [ij]}+P_{\mu \, (ij)} & = &\left( \Pi^{-1} \right)_{i}{}^{I}\left( \delta_{I}{}^{J} \partial_\mu + g A_{\mu \, I}{}^{J} \right) \Pi_{J}{}^{k} \delta_{kj}  \,.
\eea
Finally, there is a three-form antisymmetric tensor field $S_{\mu\nu\rho \, I}$, which transforms in the fundamental representation of $SO(5)_{g}$, with field strength given by
\bea
 F_{I} 
  \ \equiv \ \diff S_{I} + g A_{I}{}^{J} \wedge S_{J} \,.
\eea

Apart from the bosonic fields there are the fermionic superpartners, which we shall mention briefly here, and set to zero in the following. First, we have four gravitini $\psi_{\mu}{}^{a}$ transforming in the spinor representation of $SO(5)_c$. Secondly there are the dilatini given by sixteen spin-$\frac{1}{2}$ fields $\lambda^{a}{}_i$, which transform under $SO(5)_c$ in the spinor-vector ($\mathbf{16}$) representation.

We shall from now on set the fermionic fields to zero. The bosonic action is given by
\begin{align}
2 \mathcal{L} \ = \ & e \left\{ R + \tfrac{1}{2} m^{2}\left( T^{2} -2 T_{ij}T^{ij} \right)- P_{\mu\, ij}P^{\mu\, ij} - \tfrac{1}{2} \left( \Pi_I{}^{i}\Pi_J{}^{j} F_{\mu \nu}{}^{IJ} \right)^{2}-m^{2}\left( \left[\Pi^{-1}\right]_i{}^{I} S_{\mu \nu \rho \, I} \right)^{2}\right\} \nn\\
&- 6m \, \delta^{IJ} S_I \wedge F_J + \sqrt{3} \, \epsilon_{IJKLM} \delta^{IN}S_N \wedge F^{JK} \wedge F^{LM}+\tfrac{1}{8m}\left(2\Omega_{5}[A]-\Omega_3[A]\right) \,,
\end{align}
where $m$ is the mass parameter and $\Omega_{3}[A]$ and $\Omega_{5}[A]$ are the Chern-Simons forms of the gauge-field $A$, which are explicitly given as
\bea
  \Omega_{3}[A] & = & \epsilon^{\mu \nu \rho \sigma \alpha \beta \gamma} \, \tr \left( A_\mu F_{\nu \rho} - \frac{2}{3} A_{\mu}A_{\nu} A_{\rho} \right) \tr \left( F_{\sigma \alpha} F_{\beta \gamma} \right) \\
  \Omega_{5}[A] &= & \epsilon^{\mu \nu \rho \sigma \alpha \beta \gamma}\, \tr \left( A_\mu F_{\nu \rho}F_{\sigma \alpha} F_{\beta \gamma} - \frac{4}{5}A_\mu A_\nu A_\rho F_{\sigma \alpha} F_{\beta \gamma} -\frac{2}{5} A_\mu A_\nu F_{\rho \sigma} A_{\alpha} F_{\beta \gamma}  \right. \nn\\
  && \qquad \qquad \quad \left. + \frac{4}{5} A_\mu A_\nu A_\rho A_\sigma A_\alpha F_{\beta \gamma} - \frac{8}{35} A_\mu A_\nu A_\rho A_\sigma A_\alpha A_\beta A_\gamma \right)  \,.
\eea
From the above Lagrangian we can find the following equations of motion for the theory
\bea \label{h4eq}
\delta_{IK} \left( \Pi^{-1} \right)_i{}^{K}\left( \Pi^{-1} \right)_i{}^{J} S_J 
 & = & 
 -\frac{1}{m} * F_I 
 + \frac{1}{4 \sqrt{3} m^{2}} \epsilon_{IJKLM} * \left( F^{JK} \wedge F^{LM} \right)\,, \\
D\Big[T^{-1}_{ik} T^{-1}_{j\ell} *F^{ij}\Big]
& = &
m \, T^{-1}_{i[k} \left( *DT_{\ell]}{}^{i}  \right)
+\sqrt{3} \, \epsilon_{i_1 i_2 i_3 k \ell}\, F^{i_1 i_2}\wedge F^{i_3}\nn\\
&& 
- \frac{6}{m} F_{I_1 I_2}\wedge F^{I_1 I_2}\wedge  F_{kl} 
- 6 m^2 \, S_k\wedge S_\ell\,,\label{gaugev}\\
D\left[ \left( T^{-1} \right)_{i}{}^{k} *DT_{kj}\right] 
& = & 
2m^2 (2 T_{ik}\, T_{kj} - T_{kk}\, T_{ij}) ( *1 )
+ 4 T^{-1}_{im}\, T^{-1}_{k\ell}\, ( *F^{m\ell})\wedge F^{kj}  \nn\\
&& 
+12m^2 T_{jk}\, ( *S^k ) \wedge S^i
-\frac{1}{5} \delta_{ij} \Big[ 4 V \, ( *1 ) +4 T^{-1}_{nm} T^{-1}_{k\ell}\, (*F^{m\ell}) \wedge F^{kn}\nn\\
&& 
+ 12m^2 T_{k\ell } \, (*S^k) \wedge S^\ell \Big]\,, \label{scalarsv} \\
R_{\mu \nu} 
& = &
P_\mu{}^{ij} P_{\nu \, ij} 
+ \left(  \Pi_{I}{}^{i} \Pi_{J \, i} F_{\mu \rho}{}^{IJ} \right) \left( \Pi_{K}{}^{j} \Pi_{L \, j} F_{\nu}{}^{\rho \, KL} \right)\nn\\
&&
+ 3 m^{2} \left[ \left( \Pi^{-1} \right)_{i}{}^{I} S_{\mu\nu\rho \, I} \right] \delta^{ij}\left[  \left( \Pi^{-1} \right)_{j}{}^{J} S_{\nu}{}^{\rho \sigma}{}_{ J} \right] \nn\\
&&
-  \frac{1}{10} g_{\mu \nu} \bigg[ m^{2} \left(  T^{2} - 2 T_{ij}T^{ij}\right)
+\left( \Pi \Pi F \right)^{2} +4 m^{2} \left( \Pi^{-1} S \right)^{2} \bigg]\,.\label{einst}
\eea
Here, by $V$ we denote the scalar potential
\bea\label{Eqn:scalarpot}
V & = & \frac{1}{2} m^{2} \left[ 2 T_{ij} T_{ij} - \left( T_{ii} \right)^{2}\right] \,.
\eea
Let us remark that the scalar matrix $T_{ij}$ can be fixed to be diagonal by an $SO(5)_g$ gauge rotation. Upon doing so, this will still leave some residual gauge symmetry.

The supersymmetry conditions for the gauged supergravity theory are given by setting the supersymmetry variations of the fermionic fields to zero. In full generality these are given by (spinor indices are suppressed)
\bea
 \delta \psi_\mu & = & 
 D_\mu \epsilon + \frac{1}{20} m T \gamma_\mu \epsilon - \frac{1}{40} \left( \gamma_\mu {}^{\nu \rho} - 8 \delta_\mu {}^{\nu}\gamma^{\rho} \right) \Gamma_{ij} \epsilon \, \Pi_I{}^{i} \Pi_J{}^{j} F_{\nu \rho}{}^{IJ} \nn\\
 &&+ \frac{m}{10 \sqrt{3}} \left( \gamma_\mu {}^{\nu \rho \sigma} - \frac{9}{2} \delta_\mu {}^{\nu} \gamma^{\rho \sigma} \right) \Gamma^{i} \epsilon \,\left( \Pi^{-1} \right)_i{}^{I} S_{\nu \rho \sigma \, I} \ = \ 0\,, \\
 \delta \lambda_i & = & 
 \frac{1}{2} \gamma^{\mu} \Gamma^{j} \epsilon \, P_{\mu \, ij} +\frac{1}{2} m \left( T_{ij} - \frac{1}{5} T \delta_{ij} \right) \Gamma^{j}\epsilon + \frac{1}{16} \gamma^{\mu \nu} \left( \Gamma_{kl} \Gamma_i - \frac{1}{5} \Gamma_i \Gamma_{kl} \right) \epsilon \, \Pi_I{}^{k}\Pi_J{}^{l} F_{\mu \nu}{}^{IJ} \nn\\
 &&+ \frac{m}{20 \sqrt{3}} \gamma^{\mu \nu \rho} \left( \Gamma_i{}^{j} - 4 \delta_{i}{}^{j} \right) \epsilon \,\left( \Pi^{-1} \right)_j{}^{I} S_{\mu \nu \rho \, I} \ = \ 0\,, 
\eea
where the covariant derivative acts on the Killing spinors as
\bea
D_{\mu} \epsilon_a = \partial_\mu  \epsilon_a +\frac{1}{4} \omega_\mu{}^{mn}\gamma_{mn} \epsilon_a + \frac{1}{4} Q_{\mu \, ij} \left( \Gamma^{ij} \right)_a{}^{b} \epsilon_b \,.
\eea
Furthermore, we have used the seven-dimensional gamma matrices $\gamma_\mu$ (with Lorentzian metric) and the five-dimensional ones $\Gamma_{i}$ (with Euclidean signature), and we have denoted by $\omega_\mu{}^{mn}$ the seven-dimensional spin connection. We refer to Appendix~\ref{App:Notation} for further details on our notation and conventions as well as explicit forms for the gamma matrices.

Lastly, any supergravity solution has to satisfy the following Bianchi identities
\begin{align}
D_{[\mu} F_{\nu \rho]}{}^{ij} & \ = \ 0\,,\\
D_{[\mu} F_{\nu \rho\sigma]}{}^{i} & \ = \ 0\,,\\
D_{[\mu} \left( F^{Q} \right)_{\nu\rho]}{}^{ij} & \ = \ 0\,,\\
D_{[\mu} \left( D_{\nu]} \Pi_I{}^{i} \right)  & \ = \ 0\,.
\end{align}

Finally, the mass parameter $m$ is related to the gauge coupling by
\bea
g \ = \ 2m \, ,
\eea
and we shall employ this to remove the explicit $m$-dependence in the following,

\subsection{Uplift to eleven-dimensional M-theory}\label{Sec:Uplift}

The maximally gauged supergravity in seven-dimensions can be obtained as a consistent truncation of eleven-dimensional supergravity reduced on a four-sphere~\cite{Nastase:1999cb,Nastase:1999kf,Cvetic:1999xp}. The corresponding uplifted eleven-dimensional metric and fields are given by
\bea
 \diff s^{2}_{11} & = & \Delta^{1/3} \diff s^{2}_{7} + \frac{\Delta^{-2/3}}{g^{2}} T^{-1}_{ij} \cD \mu^{i} \cD \mu^{i} \,,
\eea
where $\mu^{i}$, with $i=1, \ldots, 5$, are constrained coordinates on $S^{4}$ satisfying $\sum_{i=1}^{5} \mu^{i} \mu^{i}=1$, and $g$ is the seven-dimensional gauge coupling as above. Furthermore, we introduced
\bea
\Delta \ = \ T_{ij} \mu^{i} \mu^{j} \,, 
\eea
as well as 
\bea
\cD \mu^{i} \ = \ \diff \mu^{i} + g A^{ij}\mu^{j}\,, \quad \text{for} \quad i \ = \ 1,\ldots 5\,.
\eea
Finally, the four-form field strength of the eleven-dimensional M-theory is given in terms of the seven-dimensional fields as
\bea
F^{(4)}_{11} & = & 
\frac{1}{g} S_{i}\wedge \cD\mu^{i} - \mu^{i} T_{ij} \left(*_{7}S^{j}\right)\nn\\
&&+\frac{\Delta^{-2}}{24 g^{3}} 
\bigg\{
- \cU \, \mu^{i_1} \, \cD \mu^{i_2}  \wedge \cD \mu^{i_3} \wedge \cD \mu^{i_4} \wedge \cD \mu^{i_5}  \nn\\
&&\qquad \qquad+ 4 \, T^{i_1}{}_{k} \mu^{k}\mu^{\ell} \, D T^{i_2}{}_{\ell}\wedge \cD \mu^{i_3} \wedge \cD \mu^{i_4} \wedge \cD \mu^{i_5}\nn\\
&&\qquad  \qquad + (6 g\Delta) \, T^{i_1}{}_{j} \mu^{j} F^{i_2 i_3} \wedge \cD \mu^{i_4} \wedge \cD \mu^{i_5}
\bigg\} \epsilon_{i_1 i_2 i_3 i_4 i_5} \,,
\eea
where we have defined
\bea
\cU & = & 2 \, T_{ij}T^{j}{}_{k} \mu^{i} \mu^{k} - \Delta \, T_{ii} \,,
\eea
and we denoted by $*_7$ the seven-dimensional Hodge star operation. 


\section{Supergravity Ansatz}\label{Sec:Ansatz}


\subsection{Ansatz}\label{Sec:Ansatzsubsec}

In this section we introduce our main ansatz, including the asymptotic behavior, for the seven-dimensional gauged supergravity renormalization group flows.\footnote{Recall that for our purposes the asymptotic conditions in the infrared and ultraviolet are considered as boundary conditions of our metric flow equations.} We will focus on the case of a K\"ahler calibrated four-cycle inside of a Calabi-Yau threefold in M-theory. As previously mentioned, a gauged seven-dimensional supergravity solution is lacking in this case. However, there are known eleven-dimensional solutions of the form AdS$_{3}\times \text{CY}_{3} \times S^{2}$~\cite{Maldacena:1997re,Gauntlett:2006ux}. As we shall see in the following, given our more general ansatz, we find some evidence suggesting that the divergence spoiling consistent IR solutions in seven-dimensional gauged supergravity might be avoided if the K\"ahler metric has non-trivial $r$-dependence. We plan on studying possible global solutions as well as more general setups in future work~\cite{Fluder:2017coass}.

The guiding principle to set up our ansatz will be to use intuition gained from known solutions~\cite{Acharya:2000mu,Gauntlett:2000ng,Gauntlett:2001jj}, as well as general arguments for M-theory geometries involving M5-branes~\cite{Gauntlett:2006ux}.

First let us recall the precise ``calibrated K\"ahler twist". We start by considering a stack of M5-branes wrapping a K\"ahler four-manifold, which has holonomy given by $U(2) \sim U(1)_{1} \times SU(2)_{2}$. In order to ensure that there are supersymmetric solutions for generic K\"ahler manifolds, we are required to introduce a (partial) topological twist.\footnote{For us this twist will only be effective as an asymptotic ultraviolet boundary condition of the holographic RG flow.} There are two ways of doing so: On the one hand one can embed the $U(1)_{1}$ subgroup of $U(2)$, or on the other hand one may embed the $SU(2)_{2}$ part inside the $SO(5)_{R}$ R-symmetry of the six-dimensional $(2,0)$ M5-branes worldvolume theory. In the former case the K\"ahler four-cycle is a calibrated cycle inside a Calabi-Yau threefold (CY$_{3}$) and in the latter case it is a calibrated cycle inside a Calabi-Yau fourfold (CY$_{4}$). In this paper we shall focus on the former case, in which the K\"ahler four-manifold is given by a calibrated four-cycle inside a CY$_{3}$.
 Thus the K\"ahler four-manifold is a holomorphic cycle calibrated by the four-form $\frac{1}{2!} J\wedge J$, where $J$ is the complex structure two-form on CY$_{3}$. The tangent bundle to the Calabi-Yau threefold restricted to the four-cycle then splits into a tangential and a normal part
\bea
T\text{CY}_{3} \big|_{\cM_{4}} \ = \ T \cM_{4} \ \oplus \ N \cM_{4} \,.
\eea
Since Calabi-Yau manifolds have vanishing first Chern-class, we find
\bea
c_{1}(T\text{CY}_{3})  \ \equiv \ 0 \ = \  c_{1}(T \cM_{4})  + c_{1}(N \cM_{4}) \,,
\eea
and one can show that $N \cM_{4}$ is intrinsic and isomorphic to the canonical bundle of $\cM_{4}$. From this, it follows that in the regime near the M5-branes, the Calabi-Yau threefold can be described by a complex line bundle over the K\"ahler four-manifold. 

Now, we are looking for solutions in the near-horizon limit. Thus, we expect that only the local geometry of the calibrated K\"ahler four-cycle inside CY$_{3}$ and its normal bundle structure enters the construction. Therefore, the original eleven-dimensional setup 
\bea
\mathbb{R}^{1,1} \times {\rm CY}_{3} \times \mathbb{R}^{3}
\eea
should now give rise to the following M-theory supergravity geometry
\bea
\mathrm{AdS}_{3}\times \left(  S_{f}^{1} \to \cM_{4}^{(IR)} \right) \times S^{2}_{R} \times I_{\theta} \,.\label{Eqn:AdS3fibS1etc}
\eea
This is only strictly true in the infrared asymptotic limit for our case. However, to formulate a sensible ansatz, it is helpful to have this intuition in mind. In equation~\eqref{Eqn:AdS3fibS1etc}, we denote by $S_{f}^{1} \to \cM_{4}$ a circle fibration over $\cM_{4}$, which is what we expect the complex line bundle to turn into in the supergravity approximation. The two-sphere $S^{2}_{R}$ will be dual to the R-symmetry of the two-dimensional superconformal field theory.\footnote{The relevant K\"ahler calibrated cycle inside CY$_{3}$ preserves $\cN=(4,0)$ supersymmetry in two dimensions.} Finally, by $I_{\theta}$ we label an interval.\footnote{More precisely, we expect the radial directions of the $\mathbb{R}^{3}$ factor and the complex line bundle to turn into the radial direction of AdS$_{3}$ and the interval $I_{\theta}$ (see for instance~\cite{Bah:2014dsa} and~\cite{Bah:2015nva} for similar statements in different setups).} Together, the $S^{1}$ factor with the sphere $S^{2}_{R}$ and the interval $I_{\theta}$ will give topologically a four-sphere. Of course the ultraviolet boundary condition of our RG flows are simply of the form AdS$_{7}\times S^{4}$, where slices of constant radius of AdS$_{7}$ are given by $\mathbb{R}^{1,1} \times \cM^{(UV)}_{4}$. We refer to Section~\ref{Sec:UpliftAnsatz} for explicit comparison of the eleven-dimensional uplift of our ansatz with the discussion here.

For the purpose of this paper and in accordance with the above picture, we shall further restrict to the seven-dimensional gauged supergravity described in Section~\ref{Sec:7dGauged}, which is a consistent truncation of M-theory on a four-sphere~\cite{Nastase:1999cb,Nastase:1999kf,Cvetic:1999xp}. The theory has an $SO(5)_{g}$ gauge symmetry corresponding to the isometry of the four-sphere. We expect to turn on gauge fields for the $SO(2)$ subgroup in $SO(5)_{g} \to SO(2)\times SO(3)$, whereas the $SO(3)$ factor, which corresponds to the R-symmetry of the two-dimensional $\cN=(4,0)$ superconformal field theory, is assumed to survive. In particular the corresponding gauge fields should then be tuned to zero in the vacuum state.\footnote{A priori, it is not clear whether they must be turned off along the full RG flow, however it is a sensible assumption, since they are turned off in the ultraviolet as well as in the infrared.}

Now, in order to set up our supergravity ansatz, let us start by looking at the precise asymptotic (boundary) conditions for the renormalization group flows in the seven-dimensional gauged supergravity. 

\paragraph{Ultraviolet.} In the UV, we expect to have a resulting metric which is asymptotically locally AdS$_{7}$ with slices of constant $r$ being of the form
\bea
\mathbb{R}^{1,1} \times \cM^{(UV)}_{4} \,,
\eea
for an arbitrary K\"ahler four-cycle $\cM_{4}^{(UV)}$. The fact that we can pick an arbitrary metric on $\cM^{(UV)}_{4}$ comes from imposing a (partial) topological twist asymptotically in the ultraviolet. The particular topological twist we are employing here (\ie,~the topological twist corresponding the K\"ahler calibration in a CY$_{3}$) can be imposed as follows: The $SO(5)_{g}$-gauge fields for the seven-dimensional gauged supergravity are specified by the spin-connection of the arbitrary K\"ahler metric on $\cM^{(UV)}_{4}$, corresponding to the fact that the theory on the M5-branes is twisted. Therefore, we decompose the gauge group as follows
\bea
SO(5)_{g} \ \to \ SO(2) \times SO(3) \,,
\eea
where we use the $SO(2)$ factor to (partially) twist the theory. This decomposition is mirrored in the eleven-dimensional M-theory setup by the division of the transverse directions to the M5-branes into tangent and normal bundles of the special holonomy manifold. From the general discussion above, we hence expect that only the $SO(2)$-gauge fields are excited in the ultraviolet. In particular, we set all the gauge fields to be vanishing apart from the component $A^{12}$, which we fix such that it cancels the spin-connection, \ie
\bea\label{Eqn:UVtwist}
\left(\left( \omega^{(UV)} \right)_\mu{}^{mn} \gamma_{mn}+ Q_\mu {}^{ij}\Gamma_{ij}\right) \, \epsilon \ = \ 0 \,,
\eea
in the asymptotic ultraviolet regime, where $\omega^{(UV)}$ is the spin-connection of the K\"ahler four-manifold $\cM^{(UV)}_{4}$ of arbitrary metric, and $Q$ is the composite gauge field. To explicitly solve equation~\eqref{Eqn:UVtwist}, we fix projection conditions for the Killing spinors, namely\footnote{Since we have yet to specify a frame, we denote the gamma matrices here by their spacetime indices. In terms of the frame in equation~\eqref{Eqn:7dframe}, and the gamma matrices in Appendix~\ref{App:Notation}, the projection conditions read
\bea
\gamma^{3} \epsilon_{a} \ = \ 0 \,, \quad \text{and} \quad
\gamma^{4} \epsilon_a \ = \ \gamma^{6} \epsilon_{a} \  = \ \ii (\Gamma_{12})_{a}{}^{b}\epsilon_{b} \,.
\eea}
\bea
 &&\gamma^{r} \epsilon_{a} \ = \ 0 \,, \label{Eqn:proj1}\\
 &&\gamma^{\bar 1} \epsilon_a \ = \ \gamma^{\bar 2} \epsilon_{a} \  = \ \ii (\Gamma_{12})_{a}{}^{b}\epsilon_{b} \,, \label{Eqn:proj2}
\eea
where $a=1,\ldots, 4$.
It is important to notice that these projection conditions are actually $\frac{1}{2}$-BPS (\ie,~we preserve half of the supersymmetries required to implement the twist). The reason we pick those projection conditions instead of the ``fully" supersymmetric ones is due to the fact that the resulting K\"ahler metric flow equations are rather restrictive~\cite{Fluder:unpublishedfullflows}, though they should be of interest in their own right. Given these projection conditions we may fix the components of the $U(1)$ gauge field $A^{12}$ in the ultraviolet by solving~\eqref{Eqn:UVtwist} asymptotically.

\paragraph{Infrared.} In the infrared, we expect that the theory is given by a metric of the form
\bea
\text{AdS}_{3}\times \cM^{(IR)}_{4} \,,
\eea
where we denote by $\cM^{(IR)}_{4}$ the four-cycle $\cM_{4}$ at the IR fixed point (\ie,~after uniformization). As explained above, we expect now that the $SO(3)$ part of the gauge symmetry corresponds to part of the R-symmetry of the dual two-dimensional superconformal field theory, and we should not have any gauge fields turned on for it in the supergravity solution. 

Given the discussion of the infrared and ultraviolet limits, we first impose that the Killing spinors $\epsilon_{a}$ surviving the projection conditions~\eqref{Eqn:proj1} and~\eqref{Eqn:proj2}, shall be preserved along the full flow. It is then natural to consider an ansatz for the seven-dimensional metric as follows
\be\label{Eqn:MetricAnsatz}
\diff s^2=e^{2 f} \diff x^2\left( \mathbb{R}^{1,1} \right) +e^{2 g} \diff r^2 + e^{2h} \diff s^2 \left( \cM_{4} \right) \,.
\ee
Here $\diff s^2 \left( \cM_{4} \right)$ is the metric on the calibrated K\"ahler four-cycle along the full RG flow, which we write as
\bea
\diff s^2 \left( \cM_{4} \right) \ = \ \left( \partial_{z_i}\partial_{\bar z_{\bar \jmath}} \cK \right) \, \diff z^{i} \diff \bar z^{\bar \jmath}\,,
\eea
where $\cK$ is the K\"ahler potential, which we pick to be an arbitrary function of $r$ as well as the coordinates on $\cM_{4}$, \ie
\bea
\cK \ \equiv \ \cK\left( r, z_1 , \bar z_{\bar 1}, z_2 , \bar z_{\bar 2} \right) \,.
\eea
The part $\diff x^2\left( \mathbb{R}^{1,1} \right)$ corresponds to the flat space metric of the resulting two-dimensional superconformal field theory. Finally, the functions $f$, $g$ and $h$ depend on the radial coordinate $r$ as well as on the holomorphic coordinates $\left\{  z_1 , \bar z_{\bar 1}, z_2 , \bar z_{\bar 2} \right\}$ of the K\"ahler four-manifold $\cM_4$, \ie
\be
 f \ \equiv \ f \left( r, z_1 , \bar z_{\bar 1}, z_2 , \bar z_{\bar 2} \right) \,, \quad 
 g \ \equiv \ g \left( r, z_1 , \bar z_{\bar 1}, z_2 , \bar z_{\bar 2} \right)   \,, \quad 
 h \ \equiv \ h \left( r, z_1 , \bar z_{\bar 1}, z_2 , \bar z_{\bar 2} \right) \,.
\ee
They have to satisfy specific asymptotic conditions in the UV and the IR, which we shall discuss in some detail in Section~\ref{Sec:Asymptotics}. 

In the following we are required to explicitly pick a frame for the seven-dimensional metric. We choose the following vielbeins for the seven-dimensional metric ansatz
\beaa\label{Eqn:7dframe}
&e^{1} \ = \ e^f \, \diff t\,, \qquad
e^{2} \ = \ e^f \, \diff x\,, \qquad
e^{3}  \ = \ e^{g}\,  \diff r \,,& \\ 
& e^{4} \ = \   e^h \, E^{1}\,,\qquad
e^{5} \ = \ e^h\, {\bar E}^{\bar 1} \,,\qquad
e^{6}  \ = \   e^h E^{2} \,,& \\ 
& e^{7}  \ = \ e^h {\bar E}^{\bar 2}\,,
&
\eeaa
where for the frame of the K\"ahler metric we define\footnote{From here on out we shall employ the following shorthand notation
\bea
 f_{r} \ \coloneqq  \  \partial_{r}  f(r, z_i,\bar z_{\bar \imath}, \ldots )\,, \quad
 f_{i \bar  \jmath} \ \coloneqq  \  \partial_{z_i} \partial_{\bar z_{\bar \jmath}} f(r,z_i,\bar z_{\bar \jmath}, \ldots )\,, \quad \text{etc} \,,
\eea
for an arbitrary function $f$ depending on variables $(r,z_i,\bar z_{\bar \jmath}, \ldots )$\,.}
\beaa\label{Eqn:Kahlerframe}
E^{1} & \ = \ \frac{ \mathcal{K}_{{1}{\bar 1}} \, \diff z_{1}+ \mathcal{K}_{{\bar 1}{2}}\, \diff z_2}{\left( \mathcal{K}_{{1}{\bar 1}} \right)^{1/2}}\,,\qquad
& {\bar E}^{\bar 1} & \ = \  \frac{  \mathcal{K}_{{1}{\bar 1}} \, \diff {\bar z}_{\bar 1} + \mathcal{K}_{{1}{\bar 2}} \, \diff {\bar z}_{\bar 2}}{\left( \mathcal{K}_{{1}{\bar 1}} \right)^{1/2}}\,, \\ 
E^{2} & \ = \ \frac{\left( \mathcal{K}_{{1}{\bar 1}} \mathcal{K}_{{2}{\bar 2}}-\mathcal{K}_{{1}{\bar 2}} \mathcal{K}_{{\bar 1}{2}} \right)^{1/2}}{\left( \mathcal{K}_{{1}{\bar 1}} \right)^{1/2}} \, \diff z_{2}\,, \qquad 
& {\bar E}^{\bar 2} & \ = \ \frac{ \left( \mathcal{K}_{{1}{\bar 1}} \mathcal{K}_{{2}{\bar 2}}-\mathcal{K}_{{1}{\bar 2}} \mathcal{K}_{{\bar 1}{2}} \right)^{1/2}}{\left( \mathcal{K}_{{1}{\bar 1}} \right)^{1/2}} \, \diff {\bar z}_{\bar 2} \,.
\eeaa
Notice that this choice of frame requires the (tangent) four-dimensional metric to be of the form
\bea
 \left( \bar g_{4} \right)_{ab} \ = \ \left( \begin{array}{cccc} \ 0 \ & \ 1 \ & \ 0 \ & \ 0 \ \\ \ 1 \ & \ 0 \ & \ 0 \ & \ 0 \ \\ \ 0 \ & \ 0 \ & \ 0 \ & \ 1 \ \\ \ 0 \ & \ 0 \ & \ 1 \ & \ 0 \ \\  \end{array} \right) \,.
\eea

Again referring to the asymptotic conditions discussed above, it is natural to turn off all but the $A^{12}$ components of the $SO(5)_{g}$ gauge fields along the full RG flow. In terms of the seven-dimensional vielbeins we may expand the field strength as
\bea\label{Eqn:F12itoframe}
 F^{12} \ = \  \frac{1}{2} \sum_{\substack{i\neq j\\ i,j\geq 3} }^{7} \cF_{ij} e^{i} \wedge e^{j} \,,
\eea
where $\left( \cF_{ij} \right)$ is anti-symmetric, and the functions $\cF_{ij}$ depend on all but the spacetime coordinates, \ie~
\bea
\cF_{ij} \ \equiv \ \cF_{ij} (r,z_{1} ,\bar{z}_{\bar 1}, z_{2}, \bar{z}_{\bar 2})\,, \quad \forall i \neq j \,.
\eea
This ansatz for the gauge fields and the metric also implies that the scalar sector of the supergravity has to satisfy reduced symmetry transformations along the full RG flow. Let us now recall that the scalar matrix $T_{ij}$ (or similarly $\Pi_{A}{}^{i}$) can be fixed to be diagonal by an $SO(5)_g$ gauge rotation.  Thus, we may fix the composite scalars to be of diagonal form, and in particular we set as an ansatz
\be\label{Eq:PiAnsatz}
\Pi_A{}^i \ = \ \diag \left( e^{3\lambda}, e^{3\lambda}, e^{-2\lambda}, e^{-2\lambda}, e^{-2 \lambda}\right)\,,
\ee
where
\be
 \lambda \ \equiv \ \lambda\left( r, z_1 , \bar z_{\bar 1}, z_2 , \bar z_{\bar 2} \right) \,.
\ee
With this choice, the composite gauge-field $Q$ is determined by the gauge-fields via 
\be \label{relQA}
Q_\mu {}^{ij}=2m \, A_\mu{}^{ij}\,.
\ee

Finally, the three-form $S^{I}$ is generically non-vanishing. However, we can trivially solve the $S$-equation of motion by setting $S^{I}=0$.

\subsection{Uplift to eleven-dimensional M-theory}\label{Sec:UpliftAnsatz}

We now briefly discuss the uplift of our seven-dimensional gauged supergravity ansatz to eleven-dimensional M-theory. We employ the general uplift formulas detailed in Section~\ref{Sec:Uplift} and first outlined in~\cite{Nastase:1999cb,Nastase:1999kf,Cvetic:1999xp}. The eleven-dimensional metric is then given by
\bea
\diff s_{11}^{2} & \ = \ & \tilde\Delta^{1/3} \diff s_{7}^{2} + \frac{\tilde\Delta^{-2/3}}{m^{2}} \left\{ e^{6\lambda} \sin^{2}\theta \left(\diff \phi + 2m A^{12}\right)^{2}  + e^{-4\lambda} \cos^{2} \theta \diff \tilde\mu^{a}\diff \tilde\mu^{a} \right\} \nn\\
&&+ \frac{e^{2\lambda} \tilde\Delta^{1/3}}{m^{2}} \diff \theta^{2} \,,\label{Eqn:upliftedmetricAnsatz}
\eea
where $\tilde\mu^{a}$, $a=1,2,3$ are constrained coordinates such that $\tilde\mu^{a}\tilde\mu^{a}= 1$, 
\bea
\tilde\Delta \ = \ e^{-6\lambda} \sin^{2}\theta + e^{4\lambda} \cos^{2}\theta\,,
\eea
and $\theta \in [0,2\pi)$. Furthermore, $\diff s_{7}^{2}$ is the seven-dimensional metric ansatz as given in~\eqref{Eqn:MetricAnsatz}. As expected from the point of view of calibrated cycles, we see that there is an $S^{1}$ fibered over the four-cycle $\cM_{4}$, which can be viewed as the unit (co-)normal bundle on the K\"ahler cycle inside the Calabi-Yau threefold. Furthermore, as expected, there is an $S^{2}$ factor corresponding to the R-symmetry, and $\theta$ gives the interval $I_{\theta}$ as required from our previous discussion around equation~\eqref{Eqn:AdS3fibS1etc}.

\subsection{Supergravity equations in Ansatz}\label{Sec:supergravityequationsinansatz}

Let us now write down the supersymmetry equations, equations of motion and Bianchi-identities given our ansatz in Section~\ref{Sec:Ansatzsubsec}. Namely, by setting $S^{I}=0$, the $S$-equation of motion
\bea \label{Eqn:seomA}
\delta_{IK} \left( \Pi^{-1} \right)_i{}^{K}\left( \Pi^{-1} \right)_i{}^{J} S_J 
 & = & 
 -\frac{1}{m} * F_I 
 + \frac{1}{4 \sqrt{3} m^{2}} \epsilon_{IJKLM} * \left( F^{JK} \wedge F^{LM} \right)\,,
\eea
is trivially satisfied. In addition, we fix a diagonal gauge for the composite scalars $\Pi_{I}{}^{i}$ as in equation~\eqref{Eq:PiAnsatz}, and thus we find (in the alternative notation) for $T_{ij}$,
\bea
T_{ij} \ = \ \diag\left( e^{-6\lambda}\, ,\, e^{-6\lambda}\, ,\, e^{4\lambda}\, ,\, e^{4\lambda}\, ,\, e^{4\lambda}\right) \,.
\eea
Then the $F$-equation of motion simplifies to
\bea\label{Eqn:feomA}
D\Big[ e^{12 \lambda} *F^{12}\Big]
& = & 0 \,.\label{gaugev}
\eea
Similarly, the $T$-equation of motion is encoded in the following single (independent) equation
\bea\label{Eqn:teomA}
 \diff  * \diff \lambda
& = & 
\left[ m^2 e^{-2\lambda} \,+\frac{2}{15}   V \,  \right] ( *1 )
+\frac{2}{5}  e^{12\lambda}\, ( *F^{1 2})\wedge F^{1 2} 
 \,,
\eea
where the scalar potential is now simply
\bea\label{Eqn:scalarpotA}
V \ = \ - \frac{3}{2} m^{2} \left[ e^{8\lambda}+ 4 e^{-2\lambda} \right] \,.
\eea
Finally, the Einstein equation (we shall use an equivalent version in different notation here)
\bea
R_{\mu\nu} & = &
 \frac{1}{4} (T^{-1})_{ij}  D_\mu  T_{jk}  (T^{-1})_{k\ell} \, D_{\nu} T_{\ell i} 
+ \frac{1}{4} (T^{-1})_{ik}(T^{-1})_{jl}F_{\mu\rho}{}^{ij}F_\nu{}^{\rho}{}^{\, kl} 
+ \frac{1}{4}T_{ij}S_{\mu\rho\sigma}{}^{i}S^{\rho\sigma}{}_\nu{}^{j}\nn\\
&&\label{Eqn:EEAlternative}
+\frac{1}{10} g_{\mu\nu} \left( 
-\frac{1}{4}(T^{-1})_{ik}(T^{-1})_{j \ell}F_{\rho\sigma}{}^{ij}F^{\rho\sigma \, k \ell}
-\frac{1}{3}T_{ij}S_{\rho \sigma \tau}{}^{i}S^{\rho \sigma \tau \, j}+ 2 V 
\right) \,,
\eea
with the scalar potential $V$ in equation~\eqref{Eqn:scalarpot} (after inserting our ansatz, $V$ is given in equation~\eqref{Eqn:scalarpotA}) can be written as
 \bea\label{Eqn:EEA}
R_{\mu\nu} & = &
30 \left(\nabla_{\mu} \lambda\right) \left( \nabla_{\nu} \lambda \right)
+ \frac{1}{2}e^{12\lambda} \left( F^{12 } \right)_{\mu\rho} \left( F^{12} \right)_\nu{}^{\rho}
-\frac{1}{20} g_{\mu\nu} \left( 
 e^{12\lambda}  \left( F^{12} \right)^{2} - 4 V 
\right) \,,
\eea
where
\bea
\left( F^{12} \right)^{2} \ \equiv \ \left( F^{12} \right)_{\mu\nu}\left( F^{12} \right)^{\mu\nu} \,.
\eea
Before we discuss the supersymmetry conditions, let us write down the only nontrivial (Abelian) Bianchi-identity
\bea\label{Eqn:BIA}
D F^{12} & \ \equiv \ & \diff\left(  F^{12} \right)  \ = \ 0 \,.
\eea

Let us now turn to the supersymmetry conditions in our ansatz. We shall not explicitly split the spinors up, since we will not explicitly need it in the remainder. The dilatini equations for $i \in \{1,2\}$ can be written as
 \bea\label{Eqn:DilatiniA1}
0 & = &  
\gamma^{\mu} \Gamma^{i} \epsilon \, \left( \nabla_{\mu} \lambda \right)
+\frac{1}{5} m \left(  e^{-6\lambda} - 3 e^{4\lambda} \right) \Gamma^{i}\epsilon 
+ \frac{1}{10} e^{6\lambda} \gamma^{\mu \nu} \Gamma_{12} \Gamma_i  \epsilon \, \left( F^{12} \right)_{\mu \nu}  \,,
\eea
and for $j \in \{3,4,5\}$ they are given by
 \bea\label{Eqn:DilatiniA2}
0 & = &  
- \gamma^{\mu} \Gamma^{j} \epsilon \, \left( \nabla_{\mu} \lambda \right)
+\frac{1}{5} m \left( e^{4\lambda} - e^{-6\lambda} \right) \Gamma^{j}\epsilon
+ \frac{1}{10} e^{6\lambda} \gamma^{\mu \nu}  \Gamma_{12} \Gamma_j \epsilon \, \left( F^{12} \right)_{\mu \nu}  \,.
\eea
Similarly the gravitini equations in our ansatz read
\bea\label{Eqn:GravitiniA}
 D_{\mu} \epsilon  & = & 
 - \frac{1}{20} m \left( 2 e^{-6\lambda}+3 e^{4\lambda} \right) \gamma_\mu \epsilon 
 + \frac{1}{20}  e^{6\lambda} \left( \gamma_\mu {}^{\nu \rho} - 8 \delta_\mu {}^{\nu}\gamma^{\rho} \right) \Gamma_{12} \epsilon \, \left( F^{12} \right)_{\nu \rho} \,,
 \eea
 where 
\bea
 D_{\mu} \epsilon_a = \partial_\mu  \epsilon_a +\frac{1}{4} \omega_\mu{}^{mn}\gamma_{mn} \epsilon_a + \frac{g}{2} \left( A^{12} \right)_{\mu} \left( \Gamma^{12} \right)_a{}^{b} \epsilon_b \,,
\eea
where we recall that we set $g=2m$. 

\subsection{Integrability}\label{Sec:integrability}

Apart from the supergravity equations described in the previous section, we will also employ what we call ``integrability". In principle one could try to solve integrability in the usual sense, \ie~use the gravitini and dilatini variation to solve schematically
\bea
[D_{\mu}, D_{\nu}] \epsilon \   \propto \  R_{\mu\nu} + \cdots \,,
\eea
where the ellipsis denote curvatures for other bundles (\emph{e.g.} gauge field strengths). However, for our purposes it is enough to do this explicitly in our ansatz/solution. The procedure goes as follows: We use the gravitini variation to solve for 
\bea\label{Eqn:int1}
&\partial_{r} \eta \ = \ \cI_{r} \eta \,, \quad 
\partial_{z_1} \eta  \ = \ \cI_{z_1} \eta\,, \quad 
\partial_{\bar z_{\bar 1}} \ = \ \cI_{\bar z_{\bar 1}} \eta \,, &\\
\label{Eqn:int1}
&\partial_{z_2} \eta \ = \ \cI_{z_2} \eta\,, \quad 
\partial_{\bar z_{\bar 2}} \eta \ = \ \cI_{\bar z_{\bar 2}} \eta \,, &
\eea
in terms of the fields in our ansatz. Here  we used $\eta$ to denote a particular component of the Killing spinor $\epsilon$, which is preserved under the aforementioned projection conditions~\eqref{Eqn:proj1} and~\eqref{Eqn:proj2}. Furthermore in $\cI_{j}$ we schematically include all the relevant fields and their derivatives that appear when solving for the left hand side. Then we take derivatives of these equations and then the ``Schwarz integrability condition" for PDEs will give us a set of equations of the form
\bea
\partial_{r z_1} \eta \ = \  \partial_{r}\left( \cI_{z_1} \eta \right) \ \equiv \ \partial_{z_1 r} \eta \ = \  \partial_{z_1}\left( \cI_{r} \eta \right) \label{Eqn:int3}
\eea
and similarly for the other pairs of variables. By plugging equations~\eqref{Eqn:int1} and~\eqref{Eqn:int1} back into~\eqref{Eqn:int3} and its cousins, we find partial differential equations purely in terms of the fields in our ansatz, independent of $\eta$ (or $\epsilon$). Furthermore, these integrability conditions will ensure that we can locally integrate to find the Killing spinors.


\section{Metric flows for K\"ahler calibrations inside $CY_3$}\label{Sec:Solution}


\subsection{K\"ahler metric flow equations}

We now sketch our solution for the ansatz discussed in Section~\ref{Sec:Ansatz} and refer to Appendix~\ref{App:Comp} for more details.

We start by defining the following combination of fields
\bea \label{Eqn:defsLGH1}
 \Lambda & = & \lambda-f  \,,\\\ 
 G & = & 4f +g \,, \label{Eqn:defsLGH2}\\
 H & = & h-f\,.\label{Eqn:defsLGH3}
\eea
From combining the gravitini equation~\eqref{Eqn:GravitiniA} and the dilatini variations~\eqref{Eqn:DilatiniA1} and~\eqref{Eqn:DilatiniA2}, we observe that these combinations of fields only depend on three out of the five variables. However, we assume in the following that they only depend on the $r$-direction\footnote{This is an assumption which helps to simplify the equations. It is rooted in the study of K\"ahler calibration flows that preserve \emph{the maximal amount} of supersymmetry~\cite{Fluder:unpublishedfullflows}. Namely, in that case, one can explicitly show that equations~\eqref{Eqn:defsLGH1}~--~\eqref{Eqn:defsLGH3} are fully general. We shall not discuss ``full" flows in the current paper.}, \ie
\bea
\Lambda \ \equiv \ \Lambda(r) \,, \quad G \ \equiv \ G(r) \,, \quad H \ \equiv H(r)\,.
\eea

Using the full range of supersymmetry equations, equations of motion, Einstein equations as well as Bianchi identities including integrability conditions given in Sections~\ref{Sec:supergravityequationsinansatz} and~\ref{Sec:integrability}, we can solve for the components of $F^{12}$. In an expansion in terms of the seven-dimensional frame coordinates -- as detailed in~\eqref{Eqn:F12itoframe} -- we have written down the resulting solutions in equations~\eqref{Eqn:Fs1}~--~\eqref{Eqn:Fs2}. Similarly, using all the aforementioned field equations we can isolate the partial derivatives of the function $f$ with respect to $r$, $z_1$ and $z_2$; the resulting solutions are provided in~\eqref{Eqn:Solfr}~--~\eqref{Eqn:Solfz2}. For our purposes, we may neglect the remaining partial derivatives of $f$ with respect to the barred coordinates. 

Furthermore, we found the following solutions for the fields $\Lambda(r)$ and $H(r)$ introduced in~\eqref{Eqn:defsLGH1} and~\eqref{Eqn:defsLGH3}
\bea
\partial_{r}\Lambda(r) & = & \frac{1}{2} m e^{G+4 \Lambda} \,,\label{Eqn:SolLambda}\\
\partial_{r}H(r) &=&\frac{1}{4} \left(\partial_r G+3 m e^{G+4 \Lambda }\right) - 
\frac{1}{4}  \partial_{r} \log\left(\mathcal{K}_{ 1 \bo } \mathcal{K}_{r \bo 2 }-\mathcal{K}_{ \bo 2 } \mathcal{K}_{r 1 \bo }\right)\,.\label{Eqn:SolH}
\eea

Having fixed all these ingredients we arrive at the following set of \emph{metric flow equations} (we again refer to Appendix~\ref{App:Comp} for more details)
\bea
 t & = & t(r) \,, \label{Eqn:Kf0}\\
 t(r) & = & s(r) \,, \label{Eqn:Kf1}\\
t (r)& = & \partial_{r} \log \left( \mathcal{K}_{ 1 \bo } \mathcal{K}_{r 2 \bt }-\mathcal{K}_{ 2 \bt } \mathcal{K}_{r 1 \bo } \right) \,, \label{Eqn:Kf2}\\
\left( \log g \right)_{r \bo} & = & 0 \,, \label{Eqn:Kf3}\\
\left( \log g\right)_{r{\bt}} & = &0 \,, \label{Eqn:Kf4} \\
\left( \log g \right)_{i \bar \jmath} \, e^{F}  &=& 
\left[\frac{\left( \log g \right)_{k \bar \ell}}{\cK_{k \bar \ell}}\, e^{F}
+m\, \partial_r \log\left( \frac{\cK_{i \bar \jmath}}{\cK_{k \bar \ell}} \right) \right]\cK_{i \bar \jmath} \,,\label{Eqn:Kf5}
\eea
where $i,k \in \{1,2\} \,,$ and $ \bar \jmath, \bar \ell \in \{\bo,\bt\}$, are arbitrary, and where we have introduced the following definitions
\bea
e^{F} & \coloneqq  & e^{G-2H+6\Lambda}\label{Eqn:Fdef}\\
t& \coloneqq  &  \partial_{r} \log\left(\mathcal{K}_{ 1 \bo } \mathcal{K}_{r \bo 2 }-\mathcal{K}_{ \bo 2 } \mathcal{K}_{r 1 \bo }\right)  \label{Eqn:tdef}\,,\\
s & \coloneqq  & \partial_r \log \left( \mathcal{K}_{{{1}{\bo}}} \mathcal{K}_{{r{1}{\bt}}}-\mathcal{K}_{{{1}{\bt}}} \mathcal{K}_{{r{1}{\bo}}} \right) \label{Eqn:sdef}\,.
\eea
In particular, the equations arising in supergravity explicitly dictate that $t$ and $s$ only depend on the $r$-coordinate. Finally, we have defined
\bea
\log g \ \coloneqq  \ \log \left(\mathcal{K}_{1 \bo} \mathcal{K}_{2 \bt}-\mathcal{K}_{1 \bt} \mathcal{K}_{\bo 2}\right)\,,
\eea
which is strictly speaking $1/2$ the logarithm of the determinant of the K\"ahler metric, and thus the Ricci tensor of the four-dimensional K\"ahler manifold reads
\bea
R_{i \bar \jmath} \ \equiv \ - \ii\, \partial_{i} \partial_{\bar\jmath} \log \det \left( g_{\mu\nu} \right) \ = \ - 2 \ii\, \left(  \log g \right)_{i \bar\jmath} \,.
\eea
There is one final equation
\bea
\frac{\left( \log g \right)_{i \bar \jmath} e^{F}}{m \, \mathcal{K}_{i \bar \jmath} }& = & \frac{1}{2}\left(\partial_rG+3 m e^{G+4 \Lambda }\right) 
+\partial_r \log \mathcal{K}_{i \bar \jmath}
-\frac{s}{2}   \,, 
\eea
for arbitrary $i \in \{ 1,2\}$ and $\bar \jmath \in \{ {\bar 1}, {\bar 2}\}$. This can be used to fix $\partial_r G(r)$ in terms of the K\"ahler potential, \ie
\bea
\partial_{r}G(r)
& = & 
-3 m e^{G+4 \Lambda }
+2\,\frac{\left( \log g \right)_{i \bar \jmath} }{\mathcal{K}_{i \bar \jmath}} \frac{e^{F}}{m} 
-2\,\partial_r \log \mathcal{K}_{i \bar \jmath}
+s \,.\label{Eqn:SolGp}
\eea
As we mention in Appendix~\ref{App:Comp}, this furnishes a complete and consistent set of equations upon taking derivatives.

\subsection{Asymptotics}\label{Sec:Asymptotics}

We shall now discuss the asymptotic behavior of the K\"ahler manifold flow equations~\eqref{Eqn:Kf0}~--~\eqref{Eqn:Kf5}. In the asymptotic ultraviolet, we expect that by implementing the appropriate twist, we can pick an arbitrary K\"ahler metric. At the infrared fixed point we should end up with a K\"ahler-Einstein metric on the K\"ahler four-manifold wrapped by the M5-branes. We shall confirm these expectations explicitly in the following.

\subsubsection{Ultraviolet}

As discussed in Section~\ref{Sec:Ansatz}, in the ultraviolet limit, the metric should be asymptotically AdS$_{7}$, with the slices at constant $r$-coordinate being of the form 
\bea
\mathbb{R}^{1,1} \times \cM^{(UV)}_{4} \,.
\eea
In particular the ultraviolet region will be in the limit $r \to 0$, and the metric will have asymptotic boundary conditions as\footnote{We use the (standard) notation: For any function $f$ of the variable $x$, $f(x) \coloneqq  o \left(g(x)\right)$ in the limit $x \to 0$, if and only if $$\lim_{x\to 0} \left| \frac{f(x)}{g(x)} \right| \ = \ 0 \,.$$}
\bea
 f \ \sim \ - \log r + o\left(1\right) \,, \quad g \ \sim \ - \log r + o\left(1\right) \,, \quad h \ \sim \ - \log r+ o\left(1\right) \,.
\eea
Similarly, the scalar $\lambda$ and the $U(1)$ gauge field $F^{12}$ have to satisfy the following boundary conditions in the UV
\bea
\left( A^{12} \right) & \ \sim \ & \frac{\ii}{4m} \left( \omega^{(UV)} \right)_{ab} J^{ab} + o\left(1\right) \,,\\
\lambda & \ \sim \ &  o\left(1\right)\,,
\eea
where $\omega^{(UV)}$ is the spin-connection purely on the four-dimensional K\"ahler manifold $\cM^{(UV)}_{4}$, and
\bea
J^{ab} \ = \ \left( \begin{array}{cccc} \ 0  \ & \  1  \ & \  0  \ & \  0 \ \\  -1   & \  0  \ & \  0  \ & \  0 \ \\ \  0  \ & \  0  \ & \  0  \ & \  1 \ \\ \  0  \ & \  0  \ & -1  & \  0 \ \end{array} \right) \,.
\eea
The former condition is precisely the asymptotic implementation of the twist~\eqref{Eqn:UVtwist}. Finally the K\"ahler potential goes as
\bea
\cK (r, z_1, \bar z_{\bar 1}, z_2 , \bar z_{\bar 2}) \ \sim \ \cK^{(UV)}_{0} (z_1, \bar z_{\bar 1}, z_2 , \bar z_{\bar 2}) + r \cK^{(UV)}_{1}(z_1, \bar z_{\bar 1}, z_2 , \bar z_{\bar 2}) + o \left( r \right) \,,
\eea
for $r \to 0$.

With this asymptotic behavior of the ansatz, the function $F(r)$ in~\eqref{Eqn:Fdef} is asymptotically given by
\bea
 e^{F} \ \sim \  \mathrm{cst} \cdot r +  o(r) 
\eea
in the $r \to 0$ ultraviolet limit. Furthermore, by including higher order terms, such as $\cK^{(UV)}_{1}$, the functions $s$ introduced in~\eqref{Eqn:sdef} and $t$ in~\eqref{Eqn:tdef} are in fact well defined and vanishing in the $r \to 0$ limit. This is important in order for the functions $f$, $g$ and $h$ to be physically sensible and well-defined. It is then straightforward to observe that our set of equations does not put any constraints on the K\"ahler metric $\cK^{(UV)}_{0}$ in the limit $r\to 0$. This confirms the expected result that we may start the RG flow with an arbitrary K\"ahler metric in the UV.

\subsubsection{Infrared}

In the infrared limit -- corresponding to $r\to \infty$ -- we expect to obtain a metric solution of the form
\bea\label{Eqn:asympIR}
\text{AdS}_{3} \times \cM_{4}^{(IR)} \,,
\eea
where $\cM_{4}^{(IR)}$ is supposed to be a  ``uniformized" version of the generic K\"ahler manifold we started with in the UV. In particular, at the infrared fixed point, the K\"ahler potential will be \emph{independent} of the radial coordinate, \ie
\bea
\partial_r \cK (r, z_1, \bar z_{\bar 1}, z_2 , \bar z_{\bar 2}) \  \equiv  \ 0 \,.
\eea
Thus, in our K\"ahler metric flow equations in~\eqref{Eqn:Kf0}~--~\eqref{Eqn:Kf5}, the remaining condition reads
\bea
\frac{(\log g)_{i\bar\jmath}}{\cK_{i\bar\jmath}} \ = \ \frac{(\log g)_{k\bar\ell}}{\cK_{k\bar\ell}}\,,
\eea
for arbitrary pairs $(i \bar\jmath), (k\bar\ell) \in \{(1\bar 1), (1 \bar 2), (2 \bar 1), (2\bar 2) \}$. In terms of the Ricci tensor and the local metric on the four-manifold, this gives
\bea
\frac{R_{i\bar\jmath}}{g_{i\bar\jmath}} \ = \ \frac{R_{k\bar\ell}}{g_{k\bar\ell}}\,.
\eea
Therefore, we conclude that $\cM_{4}^{(IR)}$ is as expected precisely K\"ahler-Einstein. The fact that the ratio is independent of the $r$-coordinate follows from~\eqref{Eqn:Kf3} and~\eqref{Eqn:Kf4}. Furthermore, its independence of the local coordinates on the K\"ahler manifold is a consequence of taking derivatives of~\eqref{Eqn:SolGp}.\footnote{There is a subtlety here in that the function $s$, defined in~\eqref{Eqn:sdef}, is not well-defined for $\partial_{r}\cK \equiv 0$.}

Now we would like to explicitly see if we can find an asymptotic (consistent) IR AdS$_{3}$ solution, by considering perturbation theory. To leading order the physical IR asymptotic behavior of the metric fields $f$ and $g$ should satisfy
\bea
 f \ \sim \ - \log r  \,, \quad g \ \sim \ - \log r \,, \label{Eqn:Asympfg}
\eea
as well as
\bea
h \ \sim \ h_{0} \ \equiv \ \mathrm{constant} \label{Eqn:Asymph} \,,
\eea
for $r\to \infty$. Thus, it follows that 
\bea
e^{F} \sim \mathrm{cst} \cdot \frac{1}{r} + o\left( \frac{1}{r} \right) \,.
\eea
Similarly, there can be corrections to the K\"ahler potential, and we fix an ansatz of the form
\bea
\cK (r, z_1, \bar z_{\bar 1}, z_2 , \bar z_{\bar 2}) \ = \ \cK^{(IR)}_{0} (z_1, \bar z_{\bar 1}, z_2 , \bar z_{\bar 2}) + \cK^{(IR)}_{1}(z_1, \bar z_{\bar 1}, z_2 , \bar z_{\bar 2})\frac{1}{r} + \cdots \,.
\eea
As mentioned above, the leading order $\cK^{(IR)}_{0}$ is then required to be K\"ahler-Einstein with some constant $\ell_{0}\in \mathbb{R}$, \ie~
\bea
\left( \cK^{(IR)}_{0} \right)_{i\bar\jmath} \ = \ \ell_{0} \left( R^{(IR)}_{0} \right)_{i \bar\jmath}\,.\label{Eqn:KEasympell0}
\eea
The subleading contributions of the K\"ahler potential can be fixed order-by-order. By doing so, we find an expansion consistent with the asymptotic behavior in~\eqref{Eqn:Asympfg} and~\eqref{Eqn:Asymph}. One can check that by solving the equations~\eqref{Eqn:SolLambda},~\eqref{Eqn:SolH} and~\eqref{Eqn:SolGp} (which determine $\Lambda$, $H$ and $G$) order-by-order, that the scalar $\lambda$ obeys
 \bea\label{Eqn:lambda0asymp}
 \lambda \ \sim \ \lambda_0 + o\left( \frac{1}{r} \right) \,,
 \eea
 where $\lambda_0$ is a finite constant. Thus, the divergence, $\lambda \to \infty$, observed in~\cite{Gauntlett:2000ng} seems to be avoided in this expansion. The main difference is that the K\"ahler potential--dependent terms in equations~\eqref{Eqn:SolH} and~\eqref{Eqn:SolGp} add more degrees of freedom, which allow us to cancel the apparent unphysical behaviour of $\lambda$. For instance, in equation~\eqref{Eqn:SolH}, the $\cK$--dependent piece starts contributing at leading order, \ie~at order $\cO\left( 1/r \right)$. If we fix the ``bulk" $\cK$ to be independent of $r$, this term will be absent, and solutions will require unphysical asymptotic behavior for the scalar $\lambda$~\cite{Gauntlett:2000ng}.\footnote{Let us stress here that we do not recover the setting in~\cite{Gauntlett:2000ng} from our set of equations~\eqref{Eqn:Kf0}~--~\eqref{Eqn:Kf5}. This is due to the fact that if the K\"ahler potential is independent of $r$ everywhere, the functions $s$ and $t$ are not well-defined (see for instance equations~\eqref{Eqn:Ass1} and~\eqref{Eqn:Ass1}).}
 
This works for the asymptotic K\"ahler-Einstein metric $\cK^{(IR)}_{0}$ being positively curved, \ie~having $\ell_{0} > 0$ in~\eqref{Eqn:KEasympell0}. However, when $\ell_{0} < 0$, we require imaginary $\lambda_{0}$ in~\eqref{Eqn:lambda0asymp}, to have a consistent set of asymptotic solutions. Of course none of this means that there are any global solutions for $\ell_{0}>0$ or $\ell_{0}<0$, and it would be desirable to derive full global solutions for either of those cases using our set of equations. We leave this for future investigation.

Finally, let us briefly dwell on the observation that the infrared metric is K\"ahler-Einstein. In the mathematics literature, K\"ahler-Ricci flows have been an active area of research for some time now (see for instance~\cite{2012arXiv1212.3653S} for a nice review). An important fact which was proved in~\cite{Cao1985} is that for non-positive first Chern-class -- and after proper normalization -- the K\"ahler-Ricci flow \emph{converges} to a K\"ahler-Einstein metrics (as a ``by-product", the author was able to re-prove the famous Calabi-Yau theorem~\cite{CPA:CPA3160310304}). However, if the first Chern class of the K\"ahler manifold is positive, the K\"ahler-Ricci flow may not converge to a K\"ahler-Einstein metric (there exist compact K\"ahler manifolds of $c_1>0$ which do not admit a K\"ahler-Einstein metric~\cite{1983InMat..73..437F,Tian1997}). It is rather interesting, that we seem to find somewhat complementary data, \ie~from a physical perspective, when $\ell_0>0$, we find evidence for ``well-behaved" IR K\"ahler-Einstein fixed points and when $\ell_{0}<0$, we have to consider a ``complex" scalar (of course, from a mathematical perspective the physical validity of the fixed points might not be relevant). The known mathematical results are suggestive that there might exist global flows in the case of $\ell_{0}<0$. Either way we believe that our set of equations could be an interesting alternative way to study four-dimensional K\"ahler manifold flows and we hope to return to that in the future. 


\section{Discussion and outlook}\label{Sec:Discussion}

In this paper we initiated the study of K\"ahler four-manifold flows by treating renormalization group flows across dimensions from holography. We started by setting up a \emph{physically sensible} ansatz for the case of M5-branes wrapping a K\"ahler four-manifold, which is a calibrated cycle inside a Calabi-Yau threefold. This ansatz is taken in the local picture of gauged seven-dimensional supergravity. We then went ahead and imposed $\frac{1}{2}$-BPS conditions on the Killing spinors, and solved all the constraints coming from the supergravity theory. This left us with a system of partial differential equations purely in terms of the four-dimensional K\"ahler metric~\eqref{Eqn:Kf0}~--~\eqref{Eqn:Kf5}. We then provided evidence that these equations should describe some sort of higher dimensional analogue of uniformization of the four-manifold, by taking expansions around the ultraviolet and infrared (physically motivated) boundary conditions. We argued that in the ultraviolet one may start with an arbitrary K\"ahler metric, and in the infrared it should uniformize to a K\"ahler-Einstein (constant curvature) metric. In particular, both the UV and IR expansions seem to be physically well-behaved.

We shall now present a rather extensive list of interesting future directions.

To begin with, an interesting problem is to study and analyze our equations~\eqref{Eqn:Kf0}~--~\eqref{Eqn:Kf5} in more detail. To do so, it might be useful to write down a ``covariantized" version of them. It is possible that this requires us to relax some assumptions in our ansatz; for instance we suspect that when we ``gauge-fix" the composite scalars to be diagonal (see~\eqref{Eq:PiAnsatz}), we also ``pick a gauge" in a possible more general ``covariantized" version of the metric flow equations, such that they reduce to the ones we found. More precisely we could imagine that fixing the composite scalars to this diagonal form might have drawn us to a specific representation of more general ``covariantized" flow equations for K\"ahler four-manifolds.

Along the same lines, it would also be very interesting to attempt to show uniformization for K\"ahler four-manifolds similar to the discussion in~\cite{Anderson:2011cz}. Given the form of our equations~\eqref{Eqn:Kf0}~--~\eqref{Eqn:Kf5}, this looks like a rather daunting task. More realistically, it would be nice to find and discuss (simple) examples of such K\"ahler flows and observe uniformization-behavior on a case-by-case basis. Similarly, it would be intriguing to analyze possible finite-time singularities, which are ubiquitous in Ricci flows, for our K\"ahler metric flow equations. If such singularities appear, there might be a way to understand them physically. Finally, it is important to find global solutions interpolating between the AdS$_{7}$ and AdS$_{3}$ boundary conditions. As we have seen, such solutions would require the K\"ahler four-cycle metric to explicitly depend on the radial coordinate, \ie~$\partial_{r} \cK_{i \bar\jmath} \neq 0$.

Another generalization to consider in the future is to treat metric RG flows for other examples of calibrated four-cycles. The case we aspire to the most would be to understand M5-branes wrapping a \emph{coassociative cycle} inside a $G_2$ manifold. This example is interesting, because for any choice of the four-manifold, the $G_2$-manifold looks locally like the bundle of self-dual two-forms over the calibrated coassociative cycle (see for instance~\cite{Acharya:2004qe}). Following the logic advertised in the current paper, and in particular the fact that we work completely locally, we would expect that starting from \emph{any} four-manifold in the ultraviolet we would get some uniformized version of the initial UV four-manifold in the infrared.\footnote{From the known solution in~\cite{Gauntlett:2000ng}, in which the metric on the four-manifold is kept fixed, we expect that in the infrared the metric is conformally half-flat, \ie~the Weyl tensor is anti-self-dual (\emph{e.g.} see~\cite{Itoh1993}).} Therefore, the answer to finding such flow equations might hint towards many interesting and largely unexplored questions in the mathematics of four-manifolds. As a matter of fact, the current paper is supposed to represent a stepping stone towards that goal. On the field theory side, one expects that the two-dimensional theory preserves $\cN=(2,0)$ supersymmetry, and a study of the two-dimensional theories and their relation to four-manifold geometry was performed in~\cite{Gadde:2013sca}.

Let us now briefly mention some observations about the coassociative four-cycle flows. To begin with, one can slightly simplify the problem by working with a generic Hermitian four-manifold instead of a fully general one. Even the case of Hermitian four-manifold flows is vastly unexplored (one of the main issues being that the Ricci flows does not seem to preserve the Hermiticity along the flow). Restricting to Hermitian four-manifolds, we are required to keep the full non-Abelian $SU(2)$ gauge fields for the twist. This will induce a large system of equations for the metric, when studying the gravitini and dilatini variation of the maximal seven-dimensional gauged supergravity. To isolate the conditions on the metric one needs to study their integrability conditions and remove all the remaining components of the $SU(2)$ gauge fields, which is very labor-intensive. Fortunately the $S$-equations of motion for the three-form in the gauged supergravity are still trivially satisfied by setting $S^{I}=0$, and so we still believe that such an approach is within the reach of possibility. 

An alternative approach is to rephrase the question inspired by intuition gained from the ``AGT -- correspondence"~\cite{Alday:2009aq}.\footnote{We thank S.~Gukov for discussing this idea with us.} Namely, instead of dealing with a seven-dimensional gauged supergravity, one truncates the supergravity ansatz to a five-dimensional theory, which ``lives" on the four-manifold together with the $r$-direction of AdS$_{7}$. This five-dimensional theory should then describe the metric RG flow. Ideally one would like to map it to a familiar five-dimensional supergravity, and then use known results, such as the study of allowed metrics on such theories to say something about the allowed metric RG flows. Furthermore, having such a theory one could hope that there are quantities in the theory that could serve as a ``$C$-function"~--~analogue along the RG flow, and possibly make a connection to the treatment in~\cite{Bobev:2017uzs}. As a matter of fact, in~\cite{Fluder:2017hrb} we employ this idea to reformulate the supergravity solutions of a particular set of four-dimensional $\cN=1$ superconformal field theories, which arise from compactifications of M5-branes on a Riemann surface~\cite{Bah:2011vv,Bah:2012dg}. We shall find that this leads to the relation between these fixed points and the study of Morse theory on two-dimensional Yang-Mills~\cite{Atiyah:1982fa}, which was originally observed in~\cite{Beem:2012yn}.

Finally, it would of course be nice to study three-manifold flow equations and AdS$_{4}$ solutions arising from a similar setup, and see if one can observe behavior akin to the Ricci flow for the metrics of the involved three-manifold. 


\section*{Acknowledgments}
\noindent
The author especially thanks Sergei Gukov for collaboration during the early stages of this project and various interesting and helpful discussions along the way. We would also like to thank Ibrahima Bah, Chris Beem, and James Sparks for insightful discussions, as well as Maxime Gabella for work on related ideas, and Nikolay Bobev and Ying-Hsuan Lin for comments on an initial draft. The work of MF is supported by the David and Ellen Lee Postdoctoral Scholarship and the U.S. Department of Energy, Office of Science, Office of High Energy Physics, under Award Number DE-SC0011632. MF thanks the Aspen Center for Physics for hospitality during the course of this work.


\appendix 


\section{Notation}\label{App:Notation}

Let us mention some notation that we employ throughout the paper. To begin with, we use the following conventions for indices:
\begin{itemize}
 \item We use capital Roman indices $I,J, \ldots \in \{ 1, \ldots, 5 \}$ to denote indices for the gauge group $SO(5)_g$. They are raised and lowered via $\delta^{AB}$.
 \item Lower letter Roman indices $i,j, \ldots  \in \{ 1, \ldots, 5 \}$ shall be used throughout as a label for the gauge group $SO(5)_c$. They are also raised and lowered via $\delta^{ij}$.
 \item Lower letter Greek indices $\mu, \nu \ldots \in \{1, \ldots, 7\}$ denote spacetime indices and are raised and lowered by $g_{\mu\nu}$. Lower letter Roman indices from the latter part of the alphabet, $m,n, \ldots \in \{1, \ldots, 7\}$ denote vielbein indices which are raised and lowered via $\eta^{mn}$ of signature $(-, +, \ldots, +)$. Notice that the time direction will always be at $\mu = 1$.
 \item We shall mostly avoid explicitly writing down spinor indices, however if we do they will be labeled by lower case Roman letters from the beginning of the alphabet, $a, b, \ldots \in \{ 1, \ldots, 4\}$.
\end{itemize}

Throughout the paper, capital letter Gamma matrices $\Gamma_{i}$ are elements in $Cliff(5,0)$ and in order to explicitly solve the supersymmetry constraints and integrability conditions, we fix a particular basis, namely
\bea
&\Gamma_{i} \ = \ - \sigma_{2} \otimes \sigma_{i} \,, \ \text{for}\ i \in \{1,2,3\} \,, \qquad \Gamma_{4}\ = \ \sigma_{1} \otimes \mathbbm{1}_{2} \,, \qquad
\Gamma_{5} \ = \ \sigma_{3} \otimes \mathbbm{1}_{2}\,.&\label{Eqn:5dGammas}
\eea
Lower case Gamma matrices $\gamma_{\mu}$ are elements in $Cliff(6,1)$. However, since there is a four-dimensional part of the metric that is given by a K\"ahler metric, we have to pick the following flat (frame) metric
\bea
 g_{mn} \ = \ \left( \begin{array}{c c c c c c c} -1 \ & \ 0 \ & \ 0 \ & \ 0 \ & \ 0 \ & \ 0 \ & \ 0 \\ 0 \ & \ 1 \ & \ 0 \ & \ 0 \ & \ 0 \ & \ 0 \ & \ 0 \\ 0 \ & \ 0 \ & \ 1 \ & \ 0 \ & \ 0 \ & \ 0 \ & \ 0 \\ 0 \ & \ 0 \ & \ 0 \ & \ 0 \ & \ 1 \ & \ 0 \ & \ 0\\ 0 \ & \ 0 \ & \ 0 \ & \ 1 \ & \ 0 \ & \ 0 \ & \ 0\\
 0 \ & \ 0 \ & \ 0 \ & \ 0 \ & \ 0 \ & \ 0 \ & \ 1\\0 \ & \ 0 \ & \ 0 \ & \ 0 \ & \ 0 \ & \ 1 \ & \ 0\\ \end{array} \right) \,.
\eea
Similarly, we have to pick gamma matrices $\gamma_{m}$, which satisfy the gamma-matrix algebra with this $g_{mn}$ , \ie
\bea
\left\{ \gamma_{m}, \gamma_{n} \right\} \ = \ 2 g_{mn} \mathbbm{1}_{8}\,.
\eea
We start by choosing a standard set of gamma matrices, $\tilde\gamma_{m}$ with respect to the usual flat seven-dimensional Euclidean  metric $\tilde g_{mn} \ \equiv \delta_{mn}$, 
\bea
&&
\tilde \gamma_{1} \ = \ \ii \tilde\gamma_{2} \cdots \tilde\gamma_{7} \,,\qquad 
\tilde \gamma_{2} \ = \ \ii \sigma_{2} \otimes \mathbbm{1}_{4} \,, \\
&&
\tilde \gamma_{2+j} \ = \ \ii \sigma_{2}\otimes \Gamma_{j}\,, \quad j = 1, \ldots, 5 \,,
\eea
where $\Gamma_{j}$ are the five-dimensional gamma matrices in~\eqref{Eqn:5dGammas}, 
and then use a transformation matrix $P_{ab}$, such that
\bea
 \tilde g_{mn} P_{p}{}^{m} P_{q}{}^{n} \ = \ g_{pq} \,.
\eea
The appropriate seven-dimensional gamma matrices are then obtained by
\bea
\gamma_{m} \ \coloneqq  \ P_{m}{}^{q} \tilde \gamma_{q} \,.
\eea

\section{The Full Solution}\label{App:Comp}

In this section we will provide some more details of the derivation of the equations~\eqref{Eqn:Kf0}~--~\eqref{Eqn:Kf5}, discussed in the main part of the paper. We will mention here that most of this rather involved computation is performed in Mathematica. As remarked in the main text, we start by making a simplifying assumption, namely we define
\bea
 \Lambda \ = \   \lambda-f \,,\quad G \ = \  4f +g \,,\quad H \ = \  h-f\,,
\eea
where the new functions $\Lambda$, $G$ and $H$ only depend on the $r$-direction
\bea
\Lambda \ \equiv \ \Lambda(r) \,, \quad G \ \equiv \ G(r) \,, \quad H \ \equiv H(r)\,.
\eea
A priori, the function $f = f(r,z_{1} ,\bar{z}_{\bar 1}, z_{2}, \bar{z}_{\bar 2})$ still depends on all the variables. Imposing the $\frac{1}{2}$-BPS projection conditions~\eqref{Eqn:proj1} and~\eqref{Eqn:proj2}, we can solve for $H(r)$ and $\Lambda (r)$ from the supergravity equations as well as integrability conditions to find\footnote{Recall that we are employing the following shorthand notation throughout the paper
\bea
 f_{r} \ \coloneqq  \  \partial_{r}  f(r, z_i,\bar z_{\bar \imath}, \ldots )\,, \quad
 f_{i \bar  \jmath} \ \coloneqq  \  \partial_{z_i} \partial_{\bar z_{\bar \jmath}} f(r,z_i,\bar z_{\bar \jmath}, \ldots )\,, \quad \text{etc} \,,
\eea
for an arbitrary function $f$ depending on variables $(r,z_i,\bar z_{\bar \jmath}, \ldots )$\,.}
\bea
\partial_{r}\Lambda(r) & \ = \ &  \frac{1}{2} m e^{G+4 \Lambda}\,, \\
\partial_{r}H(r) & \ = \ & \partial_{r} \left( -\frac{1}{4} \log \left(\mathcal{K}_{1\bar 1} \mathcal{K}_{r \bar 1 2}-\mathcal{K}_{\bar 1 2} \mathcal{K}_{r1\bar 1}\right)+\frac{1}{4} G(r) +\frac{3}{2} \Lambda (r) \right) \,.
\eea
We can integrate the latter equation to find
\bea
H(r) \ = \ -\frac{1}{4} \log \left(\mathcal{K}_{1\bar 1} \mathcal{K}_{r \bar 1 2}-\mathcal{K}_{\bar 1 2} \mathcal{K}_{r1\bar 1}\right)+\frac{1}{4} G(r) +\frac{3}{2} \Lambda (r) + \tilde h \left(z_{1},\bar{z}_{\bar 1}, z_{2}, \bar{z}_{\bar 2}\right) \,,
\eea
where $\tilde h \left(z_{1},\bar{z}_{\bar 1}, z_{2}, \bar{z}_{\bar 2}\right)$ is an arbitrary function in terms of the variables of the K\"ahler manifold. We will not require to fix this function in the following. Finally, the function $\exp G(r)$ can be fixed in terms of the K\"ahler potential as well as $\Lambda$ and $H$ as we shall see below.

We now fix the frame in~\eqref{Eqn:7dframe} and~\eqref{Eqn:Kahlerframe}.  Given the ansatz for $F^{12}$ in equation~\eqref{Eqn:F12itoframe}, 
and solving the gravitini and dilatini variations allows us to solve for the following components of $F^{12}$\footnote{\label{FN:ref1} Notice here that we already use results which we will describe below, in order to simplify these equations.}
\bea\label{Eqn:Fs1}
&& 
\cF_{34} \ = \ 5 \ii e^{-7 f-H(r)-6 \Lambda (r)}  \frac{f_{{1}}}{\left( \mathcal{K}_{{1}{\bar 1}} \right)^{1/2}} \,,\\
&&
\cF_{36} \ = \ 5 \ii e^{-7 f-H(r)-6 \Lambda (r)} \frac{ \left(f_{{2}} \mathcal{K}_{{1}{\bar 1}}-f_{{1}} \mathcal{K}_{{\bar 1}{2}}\right)}{\left( \mathcal{K}_{{1}{\bar 1}} \right)^{1/2} \left(\mathcal{K}_{{1}{\bar 1}} \mathcal{K}_{{2}{\bar 2}}-\mathcal{K}_{{1}{\bar 2}} \mathcal{K}_{{\bar 1}{2}}\right)^{1/2}}\,,\\
&&
 \cF_{45} \ = \ 
 \frac{\ii }{4}e^{-2 f-G(r)-6 \Lambda (r)}
 \bigg[
 - \left(3 e^{G(r)+4 \Lambda (r)} m+G'(r)\right) - 2 \, \partial_r  \log \cK_{1 {\bar 1}} \nn\\
&&
\qquad\qquad
+ \partial_r \log \left(  \mathcal{K}_{{1}{\bar 1}} \mathcal{K}_{r{2}{\bar 2}}-\mathcal{K}_{{2}{\bar 2}} \mathcal{K}_{r{1}{\bar 1}}  \right)
\bigg]\,, \\
&&
\cF_{46} \ = \ 0 \,,\\
&&
\cF_{47} \ = \ \frac{\ii e^{-2 f-G(r)-6 \Lambda (r)} \left(\mathcal{K}_{{1}{\bar 2}} \mathcal{K}_{r{1}{\bar 1}}-\mathcal{K}_{{1}{\bar 1}} \mathcal{K}_{r{1}{\bar 2}}\right)}{2 \mathcal{K}_{{1}{\bar 1}} \left(\mathcal{K}_{{1}{\bar 1}} \mathcal{K}_{{2}{\bar 2}}-\mathcal{K}_{{1}{\bar 2}} \mathcal{K}_{{\bar 1}{2}}\right)^{1/2}} \,,\\
&&
\cF_{56} \ = \ \frac{\ii e^{-2 f-G(r)-6 \Lambda (r)} \left(\mathcal{K}_{{1}{\bar 1}} \mathcal{K}_{r{\bar 1}{2}}-\mathcal{K}_{{\bar 1}{2}} \mathcal{K}_{r{1}{\bar 1}}\right)}{2 \mathcal{K}_{{1}{\bar 1}} \left(\mathcal{K}_{{1}{\bar 1}} \mathcal{K}_{{2}{\bar 2}}-\mathcal{K}_{{1}{\bar 2}} \mathcal{K}_{{\bar 1}{2}}\right)^{1/2}}\,, \\
&&
\cF_{67} \ = \ -\frac{\ii }{4} e^{-2 f-G(r)-6 \Lambda (r)}
\bigg[
- \left(3 e^{G(r)+4 \Lambda (r)} m+G'(r)\right)+ 
2 \, \partial_{r} \log \mathcal{K}_{{1}{\bar 1}}\nn\\
&&
\qquad \qquad 
-2 \, \partial_{r} \log  \left(\mathcal{K}_{{1}{\bar 1}} \mathcal{K}_{{2}{\bar 2}}-\mathcal{K}_{{1}{\bar 2}} \mathcal{K}_{{\bar 1}{2}}\right)
+\partial_{r} \log  \left(\mathcal{K}_{{1}{\bar 1}} \mathcal{K}_{r{2}{\bar 2}}-\mathcal{K}_{{2}{\bar 2}} \mathcal{K}_{r{1}{\bar 1}}\right) \bigg]
\,.
\eea
The remaining components of $F^{12}$, namely $\cF_{35}\,, \ \cF_{37}$ and $\cF_{57}$ can be fixed by first realizing that there cannot be any ${\bar z}_{\bar 1} {\bar z}_{\bar 2}$-component
\bea
\cF_{57} \ = \ 0 \,.
\eea
\makeatletter
\newcommand\footnoteref[1]{\protected@xdef\@thefnmark{\ref{#1}}\@footnotemark}
\makeatother
The remaining two components can be fixed by solving and combining several of the remaining equations. After a lengthy calculation, we find the rather simple solutions\footnoteref{FN:ref1}
\bea
&& \left( \cF_{35} \right)^{2} \ = \ 
- 25 \, e^{-14 f- 2 H(r)- 12 \Lambda (r)} \, \frac{  \left( f_{{\bar 1}} \right)^{2}}{\mathcal{K}_{{1}{\bar 1}}}\,,\\
&& \left( \cF_{37} \right)^{2} \ = \ 25  \, e^{-14 f- 2 H(r)- 12 \Lambda (r)} \frac{ \left(f_{{\bar 1}} \mathcal{K}_{{1}{\bar 2}}-f_{{\bar 2}} \mathcal{K}_{{1}{\bar 1}}\right)^{2}}{\mathcal{K}_{{1}{\bar 1}} \left({\mathcal{K}_{{1}{\bar 2}} \mathcal{K}_{{\bar 1}{2}}-\mathcal{K}_{{1}{\bar 1}} \mathcal{K}_{{2}{\bar 2}}}\right)} \,.\label{Eqn:Fs2}
\eea

Finally it remains to isolate the function $f$, which still depends on all the coordinates. A very lengthy and involved computation in which we use all the equations of motion, Einstein equations, integrability conditions as well as Bianchi-identities, yields again a rather simple solution\footnoteref{FN:ref1}
\bea\label{Eqn:Solfr}
&&
f_{r} \ = \  -\frac{1}{10}\bigg( 
\partial_{r} G(r) 
+ 2 m \left(e^{-10 f -6 \Lambda} + 3 e^{G(r) + 4 \Lambda(r)}\right) 
+\partial_{r} \log\left( \cK_{ \bo 2} \cK_{1 \bt} - \cK_{2 \bt} \cK_{1 \bo} \right)\nn\\
&& \qquad \qquad \qquad
-\partial_{r} \log\left( \cK_{ \bo 2} \cK_{r 1 \bo} - \cK_{1 \bo} \cK_{r \bo 2} \right)
\bigg)\,,\\
&&
f_{z_1} \ = \ 0 \,, \label{Eqn:Solfz1}\\
&&
f_{z_2} \ = \ 0 \,.\label{Eqn:Solfz2}
\eea
The remaining two components, namely $f_{{\bar z}_{\bar 1}}$ and $f_{{\bar z}_{\bar 2}}$ are unfixed, but nonzero. In principle, by determining their derivatives with respect to the other variables, and then integrating, one could fix them. However, for our purposes this is not necessary. It is however a crucial and highly nontrivial constraint that the second order derivatives satisfy the ``Schwarz integrability conditions". For instance, we can solve for $f_{r {\bar z}_{\bar 1}}$ from the supergravity constraints and it is important that this gives the same result as $\partial_{{\bar z}_{\bar 1}} (f_{r})$, where we plug in $f_{r} $ from equation~\eqref{Eqn:Solfr}. Similarly, we require this for the other cases.

Now finally we have isolated all the non-metric components of the system and we can focus on the flow equations for the K\"ahler potential. Solving all the equations of motion including integrability conditions and Bianchi identities yields a set of eight independent \emph{order-four} (\ie~maximum of four derivatives acting on the K\"ahler potential) and five ``independent"\footnote{It turns out that they are in fact not independent, but rather can be derived from the \emph{order-four} equations by taking derivatives.} \emph{order-five} equations for the K\"ahler potential.

We start by writing down the \emph{order-four} equations. To simplify the formulas, let us first write down some definitions that we also use in the main text
\bea
e^{F} & \coloneqq  & e^{G-2H+6\Lambda}\\
t &\coloneqq &  \partial_{r} \log\left(\mathcal{K}_{ 1 \bo } \mathcal{K}_{r \bo 2 }-\mathcal{K}_{ \bo 2 } \mathcal{K}_{r 1 \bo }\right) \,,\\
s &\coloneqq  &   \partial_r \log \left( \mathcal{K}_{{{1}{\bo}}} \mathcal{K}_{{r{1}{\bt}}}-\mathcal{K}_{{{1}{\bt}}} \mathcal{K}_{{r{1}{\bo}}} \right) \,.
\eea
We furthermore will use 
\bea
\log g \ \coloneqq \ \log \left(\mathcal{K}_{1 \bo} \mathcal{K}_{2 \bt}-\mathcal{K}_{1 \bt} \mathcal{K}_{\bo 2}\right)\,.
\eea
Then the first six \emph{order-four} constraints read
\bea
s
+ \partial_r \log \left[\mathcal{K}_{{r{\bo}{2}}} \right] 
 & = & 
\frac{\mathcal{K}_{{{\bo}{2}}} \left(\mathcal{K}_{{rr{1}{\bo}}} \mathcal{K}_{{r{1}{\bt}}}-\mathcal{K}_{{r{1}{\bo}}} \mathcal{K}_{{rr{1}{\bt}}}\right)}{\mathcal{K}_{{r{\bo}{2}}}  \left(\mathcal{K}_{{{1}{\bo}}} \mathcal{K}_{{r{1}{\bt}}} -\mathcal{K}_{{{1}{\bt}}} \mathcal{K}_{{r{1}{\bo}}} \right)} \label{firstorder4}  \,, \label{Eqn:order41} \\
\left( \log g \right)_{r \bo} & = & 0 \,, \\
\left( \log g\right)_{r{\bt}} & = &0 \,, \\
\frac{e^{F} \left( \log g \right)_{{\bo}{2}} }{m} 
& = & 
\frac{1}{2}\left(G'+3 m e^{G+4 \Lambda }\right)  \mathcal{K}_{{\bo}{2}} 
+\mathcal{K}_{r{\bo}{2}} 
- \frac{1}{2} s \, \mathcal{K}_{{\bo}{2}}\,, \\
\frac{e^{F} \left( \log g \right)_{{1}{\bt}} }{m} 
& = & 
\frac{1}{2}\left(G'+3 m e^{G+4 \Lambda }\right)  \mathcal{K}_{{1}{\bt}} 
+\mathcal{K}_{r{1}{\bt}} 
- \frac{1}{2} s \, \mathcal{K}_{{1}{\bt}}
\,, \\
\frac{e^{F} \left( \log g \right)_{{1}{\bo}} }{m} 
&=& 
\frac{1}{2}\left(G'+3 m e^{G+4 \Lambda }\right)  \mathcal{K}_{ 1 \bo } 
+\mathcal{K}_{r 1 \bo } 
- \frac{1}{2} s \, \mathcal{K}_{ 1 \bo }\,, \\
\frac{e^{F} \left( \log g \right)_{2 \bt} }{m} 
&=&
\frac{1}{2}\left(G'+3 m e^{G+4 \Lambda }\right)  \mathcal{K}_{2 \bt } 
+\mathcal{K}_{r 2 \bt } 
- \frac{1}{2} s \, \mathcal{K}_{2 \bt }\,.
\eea
Notice that final four equations imply
\bea
\frac{\left( \log g \right)_{i \bar \jmath}}{\cK_{i \bar \jmath}} - \frac{\left( \log g \right)_{k \bar \ell}}{\cK_{k \bar \ell}}  &=& m e^{-F} \partial_r \log\left( \frac{\cK_{i \bar \jmath}}{\cK_{k \bar \ell}} \right) \,, \ \  i,k \in \{1,2\} \,, \ \  \bar \jmath, \bar \ell \in \{\bo,\bt\} \,.
\eea
The remaining \emph{order-four} equation reads
\bea
\left[ \frac{\left(\log g\right)_{ 1 \bo }}{\mathcal{K}_{ 1 \bo } } 
+\frac{ \left(\log g\right)_{ 1 \bt }}{\mathcal{K}_{ 1 \bt } } \right]\frac{e^{F}}{m} & = & 
\left( G'+3 m e^{G+4 \Lambda}\right)
+  \partial_r \log \mathcal{K}_{1 \bo } \mathcal{K}_{ 1 \bt } \nn\\
&&
-\frac{\left( \mathcal{K}_{ 1 \bo } \mathcal{K}_{rr 2 \bt }-  \mathcal{K}_{ 2 \bt } \mathcal{K}_{rr 1 \bo } \right)}{\mathcal{K}_{ 1 \bo } \mathcal{K}_{r 2 \bt }-\mathcal{K}_{ 2 \bt } \mathcal{K}_{r 1 \bo }} \,.
\eea
Thus together with our result from above, we find that
\bea
\frac{ \mathcal{K}_{{{1}{\bt}}} \mathcal{K}_{{rr{1}{\bo}}}-\mathcal{K}_{{{1}{\bo}}} \mathcal{K}_{{rr{1}{\bt}}}}{\mathcal{K}_{{{1}{\bo}}} \mathcal{K}_{{r{1}{\bt}}}-\mathcal{K}_{{{1}{\bt}}} \mathcal{K}_{{r{1}{\bo}}}} & = & 
-\frac{\mathcal{K}_{ 1 \bo } \mathcal{K}_{rr 2 \bt }-  \mathcal{K}_{ 2 \bt } \mathcal{K}_{rr 1 \bo }}{\mathcal{K}_{ 1 \bo } \mathcal{K}_{r 2 \bt }-\mathcal{K}_{ 2 \bt } \mathcal{K}_{r 1 \bo }} \,,
\eea
and thus
\bea
s & = & \partial_{r} \log \left( \mathcal{K}_{ 1 \bo } \mathcal{K}_{r 2 \bt }-\mathcal{K}_{ 2 \bt } \mathcal{K}_{r 1 \bo } \right) \,.
\eea

Now we turn to the order-five equations. They can be written as
\bea
0 
 &=&
\frac{1}{2} \mathcal{K}_{ \bo 2 } \left(\mathcal{K}_{ 1 \bo } \mathcal{K}_{r \bo 2 }-\mathcal{K}_{ \bo 2 } \mathcal{K}_{r 1 \bo }\right){}^2 
 \Bigg(
-t ^2
+2 \, \partial_{r} t 
- 4 \frac{ \left(\mathcal{K}_{rr 1 \bo } \mathcal{K}_{r \bo 2 }-\mathcal{K}_{r 1 \bo } \mathcal{K}_{rr \bo 2 }\right)}{\mathcal{K}_{ 1 \bo } \mathcal{K}_{r \bo 2 }-\mathcal{K}_{ \bo 2 } \mathcal{K}_{r 1 \bo }}\nn\\
&& -2 G''+\left( G' \right)^2-3 m^2 e^{2 G+8 \Lambda } \Bigg)\,, \label{Eqn:order51}\\
0 
 &=& 
\mathcal{K}_{ \bo 2 } \left(\mathcal{K}_{ 1 \bo } \mathcal{K}_{r \bo 2 }-\mathcal{K}_{ \bo 2 } \mathcal{K}_{r 1 \bo }\right){}^2 \cdot \partial_{1} t 
+\mathcal{K}_{ 1 \bo } \left(\mathcal{K}_{ 1 \bo } \mathcal{K}_{r \bo 2 }-\mathcal{K}_{ \bo 2 } \mathcal{K}_{r 1 \bo }\right){}^2 \cdot \partial_{ 2} t\,, \\
0 
&=&
\left(\mathcal{K}_{ 1 \bo } \mathcal{K}_{r \bo 2 }-\mathcal{K}_{ \bo 2 } \mathcal{K}_{r 1 \bo }\right){}^2 \cdot \partial_{2} t\,, \\
0 
&=& \left(\mathcal{K}_{ 1 \bo } \mathcal{K}_{r \bo 2 }-\mathcal{K}_{ \bo 2 } \mathcal{K}_{r 1 \bo }\right){}^2 \cdot \partial_{\bo} t\,, \\
0 
&=& \left(\mathcal{K}_{ 1 \bo } \mathcal{K}_{r \bo 2 }-\mathcal{K}_{ \bo 2 } \mathcal{K}_{r 1 \bo }\right){}^2 \cdot \partial_{\bt} t \,.
\eea
Hence, assuming that
\bea
\left(\mathcal{K}_{ 1 \bo } \mathcal{K}_{r \bo 2 }-\mathcal{K}_{ \bo 2 } \mathcal{K}_{r 1 \bo }\right) & = & \mathcal{K}_{ 1 \bo }\mathcal{K}_{ \bo 2 }  \partial_r \log \left[\frac{\mathcal{K}_{\bo 2 }}{\mathcal{K}_{1 \bo }}\right] \ \neq \  0\,,\label{Eqn:Ass1}
\eea
we can integrate to find
\bea
t & \ \equiv \ & t(r) \,,
\eea
\ie~$t $ only depends on the $r$-coordinate.

Then equation~\eqref{Eqn:order51} tells us that
\bea
t^2 - 2 t' - 4 s \, \partial_r \log \mathcal{K}_{\bo  2 } & = &  -2 G''+G'^2-3 m^2 e^{2 G+8 \Lambda }-\frac{ 4 \mathcal{K}_{rr \bo  2 }}{\mathcal{K}_{ \bo  2 }} \label{finalorder51} \,,
\eea
and (if we do a different replacement)
\bea
t^2-2 t'  -4 s  \, \partial_r \log \cK_{1\bo} & = & 
- 2 G''+G'^2-3 m^2 e^{2 G+8 \Lambda }
 - \frac{4\mathcal{K}_{rr 1 \bo }}{\mathcal{K}_{ 1 \bo }} \,.\label{finalorder52}
\eea

We notice that~\eqref{Eqn:order41} can be written as
\bea
\big[s-t(r)\big] \left( \frac{\cK_{r1\bo}\cK_{\bo 2}- \cK_{r\bo 2} \cK_{1\bo}}{\cK_{r\bo 2} \cK_{1\bo}} \right) & =&0 \,,
\eea
We shall now further assume that 
\bea
\left( \frac{\cK_{r1\bo}\cK_{\bo 2}- \cK_{r\bo 2} \cK_{1\bo}}{\cK_{r\bo 2} \cK_{1\bo}} \right)  &\  \neq \ & 0\,,\label{Eqn:Ass2}
\eea
and hence
\bea
 s  \ \equiv \ s(r)  \ \equiv \  t (r) \,. \label{Eqn:sequalst}
\eea
Independently, notice that the ``Schwarz integrability conditions" for $\log g$, \ie 
\bea
\partial_1 \left( \log g \right)_{\bar 1 2} = \partial_2 \left( \log g \right)_{1 \bar 1} , \ \text{etc.}
\eea
imply that 
\bea
\partial_i s & = & 0  \,, \qquad i \in \{1, \bar 1, 2 , \bar 2\} \,.
\eea
Thus, this is consistent with~\eqref{Eqn:sequalst}.

Finally we can look at the constraints arising from $\partial_r \left( \log g  \right)_{i\bar \jmath}$. We start by noticing that we may rewrite $\left( \log g \right)_{i \bar \j}$ as
\bea
\left( \log g \right)_{i \bar \j}
&=&
\Big\{
2 H'(r)
+\partial_r \log \cK_{i \bar \jmath} \Big\} \left(  m e^{-F}  \mathcal{K}_{i \bar \jmath} \right) \,.
\eea
Therefore, we obtain
\bea
 \partial_r \left( \log g \right)_{i \bar \j} & = & 0 \nn\\
& = &
 \partial_{rr}
\Big[
 \frac{1}{2} \left(G+6 \Lambda\right)
+\log \cK_{i \bar \jmath} 
- \frac{1}{2} \log \left( \mathcal{K}_{{{1}{\bo}}} \mathcal{K}_{{r{1}{\bt}}}-\mathcal{K}_{{{1}{\bt}}} \mathcal{K}_{{r{1}{\bo}}} \right)\Big] \left(  m e^{-G+2 H-6 \Lambda }  \mathcal{K}_{i \bar \jmath} \right) \nn\\
&&+\partial_r
\Big[
 \frac{1}{2} \left(G+6 \Lambda\right)
+\log \cK_{i \bar \jmath} 
- \frac{1}{2} \log \left( \mathcal{K}_{{{1}{\bo}}} \mathcal{K}_{{r{1}{\bt}}}-\mathcal{K}_{{{1}{\bt}}} \mathcal{K}_{{r{1}{\bo}}} \right)\Big] \partial_r \left(  m e^{-G+2 H-6 \Lambda }  \mathcal{K}_{i \bar \jmath} \right) \,. \nn\\
\eea
However, we also have
\bea
\partial_r \left(  m e^{-G+2 H-6 \Lambda }  \mathcal{K}_{i \bar \jmath} \right)  
 &=&\left[ - \frac{1}{2} \left( G'+3 m e^{G+4 \Lambda}+t(r)\right)  + \partial_r \log \mathcal{K}_{i \bar \jmath}\right] \mathcal{K}_{i \bar \jmath} m e^{-G+2 H-6 \Lambda }\,,\nn\\
\eea
as well as
\bea
\Lambda'' & = &
\frac{1}{2} m (G' + 2 m e^{G+4 \Lambda}) e^{G +4 \Lambda} \,.
\eea
Hence, we obtain
\bea
 0 & = &
 \frac{1}{4}\bigg( 2 G''- G'^{2}+ 3 m^{2} e^{2G+8 \Lambda}  
+4\partial_{rr} \log \cK_{i \bar \jmath} 
- 2\left(  t+s \right) \left( \partial_r\log \cK_{i \bar \jmath}  \right)
+4 \left(  \partial_r \log \mathcal{K}_{i \bar \jmath} \right)^{2}  \nn\\
&& \qquad
- 2 \partial_{r} s
+ t s+ \left( G'+3 m e^{G+4 \Lambda} \right) \left( s-t \right)\bigg)\,,
\eea
and so it follows that
\bea
4\partial_{rr} \log \cK_{i \bar \jmath} 
&=&
- 2 G''+ G'^{2}- 3 m^{2} e^{2G+8 \Lambda}  
+2\left(  t+s \right) \left( \partial_r\log \cK_{i \bar \jmath}  \right)
-4 \left(  \partial_r \log \mathcal{K}_{i \bar \jmath} \right)^{2}  \nn\\
&& 
+ 2 \partial_{r} s
- t s
- \left( G'+3 m e^{G+4 \Lambda} \right) \left( s-t \right) \,.
\eea
However given~\eqref{Eqn:sequalst}, we may write this as
\bea
t^2-2 t'
-4 t \left( \partial_r\log \cK_{i \bar \jmath}  \right)
&=&
- 2 G''+ G'^{2}- 3 m^{2} e^{2G+8 \Lambda}  
-4 \frac{\cK_{rr i \bar\jmath}}{\cK_{i \bar\jmath}}  \,.
\eea
Now if we pick $(i \bar\jmath) = (\bo 2)$ or $(i \bar\jmath) = (1 \bo)$, this is precisely what we found in~\eqref{finalorder51} and~\eqref{finalorder52}. 

There are two remaining choices for $(i \bar\jmath)$ arising from this equation. The fact that they are implied by the previous equations follows from the following: Let $(k, {\bar \ell})$ and $(i {\bar \jmath})$ be arbitrary, for consistency we require that
\bea
&& 
s^{2}-2 \partial_{r} s - 4s \left( \partial_r\log \cK_{i \bar \jmath}  \right)
- \bigg[s^{2}-2 \partial_{r} s - 4s\left( \partial_r\log \cK_{k \bar \ell}  \right)\bigg]\nn\\
&=&
- 2 G''+ G'^{2}- 3 m^{2} e^{2G+8 \Lambda} -4\partial_{rr} \log \cK_{i \bar \jmath} -4 \left(  \partial_r \log \mathcal{K}_{i \bar \jmath} \right)^{2}  \nn\\
&&+ \frac{ 2 \cK_{1 \bo}}{ \left(\mathcal{K}_{r 1  \bo }\mathcal{K}_{ \bo  2 }- \mathcal{K}_{ r \bo  2 }\mathcal{K}_{1  \bo }\right)} \left[\frac{1}{2} \left( G'+3 m e^{G+4 \Lambda} \right) -\frac{1}{2} s + \left( \partial_r\log \cK_{i \bar \jmath}  \right) \right]\nn\\
&&- \bigg[- 2 G''+ G'^{2}- 3 m^{2} e^{2G+8 \Lambda} -4\partial_{rr} \log \cK_{k \bar \ell} -4 \left(  \partial_r \log \mathcal{K}_{k \bar \ell} \right)^{2}   \nn\\
&&+ \frac{ 2 \cK_{1 \bo}}{ \left(\mathcal{K}_{r 1  \bo }\mathcal{K}_{ \bo  2 }- \mathcal{K}_{ r \bo  2 }\mathcal{K}_{1  \bo }\right)} \left[\frac{1}{2} \left( G'+3 m e^{G+4 \Lambda} \right) -\frac{1}{2} s + \left( \partial_r\log \cK_{k \bar \ell}  \right) \right]\bigg]\,,
\eea
which can be rewritten as
\bea
\frac{1}{2} \left[s+t\right]\left[\left( \partial_r\log \cK_{k \bar \ell}  \right) - \left( \partial_r\log \cK_{i \bar \jmath}  \right)\right]
 & = & 
\partial_{rr} \log \cK_{k \bar \ell} -4\partial_{rr} \log \cK_{i \bar \jmath} \nn\\
&&
+ \left[ \left( \partial_r\log \cK_{k \bar \ell}  \right)^{2} - \left(\partial_r\log \cK_{i \bar \jmath}  \right)^{2}\right] \,.
\eea
However, this is equivalent to
\bea
\partial_r \left( \log g \right)_{i \bar \j} -\partial_r \left( \log g \right)_{k \bar \ell} \ = \ 0 & = &  
\partial_{rr}  \log\left( \frac{\cK_{i \bar \jmath}}{\cK_{k \bar \ell}} \right)
- \frac{1}{2}\left(  t+s \right) \partial_r\log \left( \frac{\cK_{i \bar \jmath}}{\cK_{k \bar \ell}} \right)\nn\\
&&
+\left(  \partial_r \log \mathcal{K}_{i \bar \jmath} \right)^{2}-\left(  \partial_r \log \mathcal{K}_{k \bar \ell} \right)^{2} \,,
\eea
and thus we showed that this is actually implied by our \emph{order-four} equations.

\subsection{Summary}

Let us briefly summarize the independent metric flow equations arising from the analysis of the solutions. The \emph{order-four} constraints can be summarized to the following set of equations
\bea
t & = & s \,,\label{Eqn:Summ1} \\
t & = & \partial_{r} \log \left( \mathcal{K}_{ 1 \bo } \mathcal{K}_{r 2 \bt }-\mathcal{K}_{ 2 \bt } \mathcal{K}_{r 1 \bo } \right) \,, \label{Eqn:Summ11}\\
\left( \log g \right)_{r \bo} & = & 0 \,, \label{Eqn:Summ2}\\
\left( \log g\right)_{r{\bt}} & = &0 \,,\label{Eqn:Summ3} \\
\frac{\left( \log g \right)_{i \bar \jmath} }{\mathcal{K}_{i \bar \jmath}}& = & \left\{\frac{1}{2}\left(G'+3 m e^{G+4 \Lambda }\right) 
+\partial_r \log \mathcal{K}_{i \bar \jmath}
-\frac{s}{2}  \right\} m e^{-F} \,,\label{Eqn:Summ4}\\
\left( \frac{\left( \log g \right)_{i \bar \jmath}}{\cK_{i \bar \jmath}} 
-
 \frac{\left( \log g \right)_{k \bar \ell}}{\cK_{k \bar \ell}} \right)
&=& 
\partial_r \log\left( \frac{\cK_{i \bar \jmath}}{\cK_{k \bar \ell}} \right) m e^{-F} \,, \quad i,k \in \{1,2\} \,, \ \bar \jmath, \bar \ell \in \{\bo,\bt\} \,.\nn\\
\label{Eqn:Summ5}
\eea
The \emph{order-five} constraints then imply that
\bea
 && s \ \equiv  \ s (r)\,,
\eea
as well as (independently from \emph{order-four})
\bea
 && t \ \equiv  \ t (r) \,.
\eea

The logic should be as follows. We solve~\eqref{Eqn:Summ4} in terms of $\partial_{r}G$ and $e^{F}$ for different choices of $(i \bar \jmath)$. In particular for fixed $(i \bar \jmath)$ and $(k \bar \ell)$ we can solve
\bea
G'
& = & 
-3 m e^{G+4 \Lambda }
+2\,\frac{\left( \log g \right)_{i \bar \jmath} }{\mathcal{K}_{i \bar \jmath}} \frac{e^{F}}{m} 
-2\,\partial_r \log \mathcal{K}_{i \bar \jmath}
+s \,.
\eea
Plugging this into~\eqref{Eqn:Summ4} removes that equation and leaves only~\eqref{Eqn:Summ5}. We can solve this now for $e^{F}$, to find
\bea
m e^{-F}
& = &
\frac{\left( \frac{\left( \log g \right)_{i \bar \jmath}}{\cK_{i \bar \jmath}} 
-
 \frac{\left( \log g \right)_{k \bar \ell}}{\cK_{k \bar \ell}} \right)}{\partial_r \log\left( \frac{\cK_{i \bar \jmath}}{\cK_{k \bar \ell}} \right) }
 \,.
\eea
Our analysis shows that doing this is in fact consistent, \ie~
\bea
\partial_{r} e^{-F} \ = \ (-G' + 2 H' - 6 \Lambda' )e^{-F} \,, \quad \partial_{z_1} e^{-F} \ = \ 0 \,, \ \text{etc.} \,.
\eea
Once we eliminate these functions, we find equations purely in terms of the K\"ahler potential.

Let us also mention here that equations~\eqref{Eqn:Summ1} and~\eqref{Eqn:Summ11} imply that the remaining quantities are also equal, namely
\bea
s \ \equiv \ t & = & \partial_{r} \log \left( \mathcal{K}_{ 1 \bt } \mathcal{K}_{r 2 \bt }-\mathcal{K}_{ 2 \bt } \mathcal{K}_{r 1 \bt } \right)\\
& = & \partial_{r} \log \left( \mathcal{K}_{ 1 \bt } \mathcal{K}_{r \bo 2 }-\mathcal{K}_{ \bo 2} \mathcal{K}_{r 1 \bt } \right)\\
& = &\partial_{r} \log \left( \mathcal{K}_{ \bo 2 } \mathcal{K}_{r 2 \bt }-\mathcal{K}_{ 2 \bt } \mathcal{K}_{r \bo 2 }  \right)\,.
\eea
Finally, it is noteworthy that these functions might not be well-defined (\emph{e.g.} when $\partial_{r}\cK \ \equiv \ 0$). In that case however we expect to get back to the solutions discussed in~\cite{Gauntlett:2000ng} (and one can explicitly check that). It is the explicit purpose of the current paper to move away from this case, and thus our equations correspond to a disjoint class of solutions. 


\bibliography{refs}

\end{document}